\documentclass[twocolumn,epjc3]{svjour3}
\usepackage{amssymb}
\usepackage{amsmath}
\usepackage{graphicx,subfigure,color,dcolumn,booktabs,bm}
\usepackage{longtable,lscape}
\usepackage{txfonts}
\usepackage{overpic}
\usepackage{indentfirst}
\usepackage{cases}
\usepackage{multirow}
\usepackage{ulem}
\usepackage[colorlinks,
            citecolor=blue,
            anchorcolor=red,
            menucolor=red,
            linkcolor=red,
            filecolor=red,
            runcolor=red,
            urlcolor=blue,
            frenchlinks=false]{hyperref}

\graphicspath{{Figures/}}
\allowdisplaybreaks
\journalname{Euro. Phys. J. C}

\begin{document}

\title{Resonant Singly Heavy Pentaquarks in the MIT Bag Model: Mass Spectra and Strong Decays}
\author{Wen-Nian Liu\thanksref{e1,addr1,addr2}, Wen-Xuan Zhang\thanksref{e2,addr3},
Cheng-Jie Wang\thanksref{e3,addr1,addr4},
Kai-Kai Zhang\thanksref{e4,addr1},
Fu-Quan Dou\thanksref{e5,addr1,addr4}}

\date{Received: date / Accepted: date}
\thankstext{e1}{e-mail: 202411170250@nwnu.edu.cn} 
\thankstext{e2}{e-mail:zhangwx89@outlook.com}
\thankstext{e3}{e-mail:wangchengjie1007@163.com}
\thankstext{e4}{e-mail:zhangkk314@outlook.com}
\thankstext{e5}{e-mail:doufq@nwnu.edu.cn (corresponding author)}
\institute{College of Physics and Electronic
	Engineering, Northwest Normal University, Lanzhou 730070, China \label{addr1}
\and Xinjiang Laboratory Phase Transitions and Microstructures in Condensed Matters,College of Physical Science and Technology, Yili Normal University, Yining, Xinjiang,835000,China \label{addr2}
\and School of Physical Science and Technology, Lanzhou University, Lanzhou 730000, China \label{addr3}
\and Gansu Provincial Research Center for Basic Disciplines of Quantum Physics, Lanzhou 730000, China \label{addr4}
}

\maketitle

\abstract{
Exploring the limits of color interactions in multiquark states is an important topic. Based on the bag confinement picture of hadrons, we find that for singly heavy pentaquarks, the bag confinement radius precisely falls within the range of color interaction limits provided by lattice QCD, approximately $1.17--1.29 \, \text{fm}$. This leads us to believe that singly heavy pentaquark states have the potential to form resonant states. Inspired by singly heavy baryons, we consider the mirror pentaquarks of singly heavy baryons. Furthermore, we adopt the MIT bag model, taking into account chromomagnetic and color-electric interactions between heavy and strange quarks, to calculate the mass spectrum of singly heavy pentaquarks configured as $qqqQ\bar{q}$
and analyze the stability of their S-wave two-body strong decays. We show that for the singly heavy pentaquark system, the masses are generally about $500\, \text{MeV}$ higher than the corresponding mirror baryon ground state masses, which is consistent with conclusions drawn from chiral methods. We also provide a mass mapping relationship between singly heavy pentaquarks and singly heavy baryons based on light quark flavor symmetry. The analysis of strong decays indicates that these singly heavy pentaquarks are unstable with respect to strong decays, which is consistent with our initial hypothesis.

\section{Introduction}

At the early stages of the quark model, the concept of exotic hadrons, such as tetraquarks and pentaquarks, was introduced \cite{Gell-Mann:1964ewy,Zweig:1964ruk}, which sparked theoretical and experimental investigations into these multiquark states. In 2003, the Belle collaboration first reported the candidate for the hidden charm tetraquark state $X(3872)$ \cite{Belle:2003nnu}. In the following years, a number of exotic hadrons with hidden charm or double charm tetraquark components were reported, including the well-known $Z_c(3900)$ \cite{BESIII:2013ris} and $T_{cc}(3875)$ \cite{LHCb:2021auc}, as well as the recently discovered $X(6900)$ \cite{LHCb:2020bwg} and $X(6600)$ \cite{CMS:2023owd}. Meanwhile, the LHCb collaboration revealed several signals of pentaquark components in the $J/\psi p$ and $\bar{p}$
channels, such as $P_c(4380)^+$ \cite{PhysRevLett.115.072001}, $P_c(4457)^+$, $P_c(4440)^+$, $P_c(4312)^+$ \cite{LHCb:2019kea}, and $P_c(4337)$ \cite{LHCb:2021chn}. More recently, the LHCb collaboration discovered pentaquark states with strangeness $S = -1$, namely $P_{cs}(4459)^0$ \cite{LHCb:2020jpq} and $P_{\psi s}^{\Lambda}(4338)^0$
\cite{LHCb:2022ogu} in the $J/\psi\Lambda$ channel. Theoretical works have also delved into the structures of these exotic hadrons in greater detail, primarily focusing on hadronic molecular states or compact multiquark states \cite{Chen:2022asf,Liu:2019zoy,Chen:2016qju,Karliner:2021xnq,Wang:2019nvm,Zou:2021sha,Deng:2022vkv,Liu:2024uxn}. With the substantial experimental results accumulated for double charm exotic hadron states, we are increasingly hopeful for the potential discovery of pentaquarks in the singly heavy sector \cite{ParticleDataGroup:2024cfk}.

A natural idea is to consider inserting a quark pair $n\bar{n}$
into hadrons, which is referred to as the pentaquarks mirror of baryons. Based on chiral symmetry, it was proposed that the negative-parity baryon $N^{\ast}(1535)$ could serve as a pentaquark mirror of baryons \cite{PhysRevD.39.2805,10.1143/PTP.106.873,PhysRevD.82.014004,PhysRevD.99.034012}, a notion supported by lattice QCD \cite{PhysRevLett.59.399}. In Ref. \cite{PhysRevD.104.034009}, this symmetry was extended to the study of excited states of singly heavy baryons, supporting the notion that the states $\Lambda_{c}(2765)$
and $\Xi_{c}(2967)$
are predominantly pentaquarks, while also predicting the presence of negative-parity singly heavy pentaquark components. In fact, the chiral model allows for the generation of quark pairs $n\bar{n}$
from the vacuum in the excited states of mesons or heavy baryons when their mass exceeds that of the ground state by approximately $500 \text{ MeV}$ \cite{Arifi:2020yfp}.

When the quark model was first proposed, Jaffe and others introduced the classical MIT bag model \cite{Jaffe:1976ig,Jaffe:1976ih,DeGrand:1975cf}, which was initially used to study multi-quark states \cite{Barnes:1982tx,Chanowitz:1982qj,Strottman:1979qu}. In Ref. \cite{Zhang:2021yul}, the chromomagnetic interaction (CMI) between quarks and the color-electric interactions of heavy quarks were considered, resulting in a good fit to the mass spectra, magnetic moments, and charge radii of mesons and baryons. This model has also been extended to calculations of multiquark states, such as the fully heavy tetraquark state $X(6600)$ \cite{Zhang:2023hmg}, hidden charm tetraquarks \cite{Yan:2023lvm}, triply heavy tetraquarks \cite{Zhu:2023lbx}, as well as discussions on the hidden charm pentaquark states $P_{cs}(4338)$
and its spin partner $P_{cs}(4459)$
\cite{Zhang:2023teh}. Recent studies on the bag model have sparked discussions regarding the bag binding term and its stability \cite{Liu:2025fbe}. The bag confinement condition requires that hadrons become unstable when the bag radius exceeds $1.1\,\text{fm}$, which is consistent with findings from lattice QCD and the Schwinger model, indicating that the QCD vacuum is unstable in the range of $1.17$ to $1.29\,\text{fm}$, leading to string breaking \cite{BALI20011,Bulava:2019iut,Bali:2005fu,Kou:2024dml}. The numerical results from both lattice QCD and the bag model are relatively consistent, although the bag model provides a more conservative confinement limit. Compared to lattice QCD, the bag model offers a hadronic perspective on understanding the confinement mechanism, making it easier to extend bag confinement to the study of multiquark states.

In the present work, we systematically investigate the singly heavy pentaquark system based on the bag model. We find that the average bag radius for the $nnnQ\bar{n}$ system lies in the range of $1.16\textendash1.19\, \text{fm}$. This scale is slightly higher than the limiting scale of the bag model at $1.11\, \text{fm}$, but lower than the minimum radius estimated for molecular states within the bag model. Notably, this scale corresponds precisely to the range of interaction limits provided by lattice QCD, leading us to estimate that singly heavy pentaquarks exhibit strong resonant properties. Additionally, we consider the chromomagnetic interaction between quarks and the color-electric interactions between heavy quarks and strange quarks, calculating the mass spectra for strangeness $S=0, -1, -2$. Our results show that for most of the ground state pentaquarks, the mass exceeds the corresponding mirror baryon mass by approximately $500\, \text{MeV}$, and all states have masses above the lowest thresholds for baryon-meson pairs. Additionally, we analyze the strong decay stability of various final state representations for singly heavy flavor pentaquarks, revealing that most of these states are strongly decay unstable, although a few states exhibit narrow widths due to threshold effects.

The remainder of this paper is organized as follows: In Section \ref{MIT bagmodel}, we introduce the MIT bag model and its related parameters. In Section \ref{QCD limit}, we analyze the confinement form of the bag model and the bag radius of the $nnnQ\bar{n}$
system. Section \ref{sec:hadrons} discusses the mass spectrum of the $nnnQ\bar{n}$ system and the stability of strong decays. Finally, in Section \ref{sec:summary}, we provide a brief summary.

\section{Method for MIT bag model}\label{MIT bagmodel}

In the bag picture, quarks are confined within a spherical bag, with light quarks moving at relativistic speeds inside the bag while heavy quarks occupy the center of the bag and recoil \cite{Jaffe:1976ig,Jaffe:1976ih,DeGrand:1975cf}. Taking into account the electromagnetic interactions between quarks and the color electric interactions of heavy quarks, the mass formula of the bag model is given by

\begin{equation}
	M\left( R\right) =\sum_{i}\omega _{i}+\frac{4}{3}\pi R^{3}B-\frac{Z_{0}}{R}%
	+H_{BE}+H_{CMI},\label{M}
\end{equation}%
where 
\begin{equation}
	\omega _{i}=\left( m_{i}^{2}+\frac{x_{i}^{2}}{R^{2}}\right) ^{1/2},
	\label{freq}
\end{equation}%
here, the first term on the right side represents the sum of the relativistic energies of the quarks, while the second term describes the volume energy, with the parameter $B$ understood as the energy density within the hadron's bag. The third term corresponds to the vacuum zero-point energy, with $Z_{0}$ being a bag parameter. The terms $H_{BE}$ and $H_{CMI}$ describe the CMI between quarks and the color-electric interactions between heavy quarks, respectively. The bag radius $R$ is the only variable in Eq. (\ref{M}), which can be determined using the variational method. In Eq. (\ref{freq}), $x_{i}$ represents the momentum of the quarks, satisfying the boundary condition of the bag surface, which takes the form of a transcendental equation

\begin{equation}
	\tan x_{i}=\frac{x_{i}}{1-m_{i}R-\left(
		m_{i}^{2}R^{2}+x_{i}^{2}\right)
		^{1/2}}.  \label{transc}
\end{equation}

We consider the short-range interactions between heavy quarks, denoted as $H_{BE}$. The Refs \cite{Karliner:2014gca,Karliner:2017elp,Karliner:2017qjm,Karliner:2020vsi} suggests that these short-range interactions are fundamentally color-electric interactions among heavy quarks. Specifically, this includes short-range interactions between pairs such as $c-s$ and $b-s$, as well as interactions among other heavy flavor quarks. Based on Refs. \cite{Zhang:2021yul,Yan:2023lvm}, we present the form of the color-electric interactions between heavy quarks as
\begin{equation}
	\ H_{BE} = \sum_{i<j} (\lambda_i \cdot \lambda_j) \, \frac{\alpha_s(R)}{R} E_{ij}, 
\end{equation}
on the right-hand side of the equation, $\lambda_i \cdot \lambda_j$ represents the color factor, $\alpha_s(R)$ denotes the running coupling constant, and $E_{ij}$ refers to the color-electric interactions between heavy quarks or between heavy quarks and strange quarks. The value of $E_{ij}$ for the color $\boldsymbol{\bar{3}}_{c}$ configuration can be determined by fitting experimental data from heavy-flavor mesons \cite{Zhang:2021yul}:

\begin{equation}
	\begin{Bmatrix}
		E_{cs}=-0.025\,\text{GeV}, & E_{bs}=-0.032\,\text{GeV}. \\
	\end{Bmatrix}\label{equ:binding}
\end{equation}
The short-range interactions described above can be generalized to multi-quark configurations through the color factor.

In this formulation, we incorporate the CMI among confined quarks within the bag model framework. The corresponding interaction Hamiltonian takes the form:
\begin{equation}
	\begin{aligned}
		H_{CMI}=-\sum_{i<j}^{}(\lambda_{i}\cdot \lambda_{j})(\sigma_{i} \cdot \sigma_{j})C_{ij}. 
	\end{aligned}
	\label{eq:eq6}    
\end{equation}
Within the CMI framework, the indices \( i \) and \( j \) label constituent quarks/antiquarks, while \( \lambda_i\) (Gell-Mann matrices) and \( \sigma_j \) (Pauli matrices) encode $\textbf{SU(3)}_{c}$ and $\textbf{SU(2)}_{s}$ symmetries, respectively. The coefficients \( C_{ij} \) quantify pairwise interaction strengths. For explicit evaluation of color-spinor matrix elements, we adopt the following operator representation:

\begin{equation}
	\begin{aligned}
		\left\langle\lambda_{i}\cdot\lambda_{j}\right\rangle_{nm}=
		\sum_{\alpha=1}^{8}\mathrm{Tr}(c_{in}^{\dagger}\lambda^{\alpha}c_{im})
		\mathrm{Tr}(c_{jn}^{\dagger}\lambda^{\alpha}c_{jm}),
	\end{aligned}
	\label{eq:eq7}    
\end{equation}

\begin{equation}
	\begin{aligned}
		\left\langle\sigma_{i}\cdot\sigma_{j}\right\rangle_{xy}=\sum_{\alpha=1}^{3}
		\mathrm{Tr}(\chi _{ix}^{\dagger}\sigma ^{\alpha}\chi_{iy})\mathrm{Tr}(\chi _{jx}^{\dagger}\sigma^{\alpha}\chi _{jy}),
	\end{aligned}
	\label{eq:eq8}    
\end{equation}
the expression for the coupling constant is given by
\begin{equation}
	\begin{aligned}
		C_{ij}=3\frac{\alpha_{s}(R)}{R^{3}}\bar{\mu}_{i}\bar{\mu}_{j}I_{ij}.
	\end{aligned}
	\label{eq:eq9}    
\end{equation}

In the MIT bag model \cite{Wang:2016dzu}, after performing a variational calculation for the mass, the quark spinor wave function $\psi_{i}(r)$ that satisfies the boundary conditions can be expressed as follows:

\begin{equation}
	\psi_{i}(r) = N_{i} \binom{j_{0}(x_{i}r/R)U}
	{i\frac{x_{i}}{(\omega_{i}+m_{i})R}j_{1}(x_{i}r/R)\boldsymbol{\sigma}\cdot\boldsymbol{\hat{r}}U} e^{-i\omega_{i}t}. \label{equ:quarkspinor}
\end{equation}
Therefore, the quark magnetic moment $\mu_{i}$
here is not fixed as a parameter, but rather is the expectation value of the operator $\boldsymbol{r \times \gamma}$ acting on the spinor wave function $\psi_{i}(r)$. Furthermore, we express the quark magnetic moment as follows:
\begin{equation}
	\begin{aligned}
		\mu_{i} &= \frac{Q_{i}}{2} \int_{bag}\mathrm{d}^{3}r\, 
		\bar{\psi_{i}}\left(\boldsymbol{r\times\gamma}\right)\psi_{i} \\
		&= \frac{Q_{i}}{2} \int_{0}^{R}\mathrm{d}r\,r^{2} \int\mathrm{d}\Omega\
		\bar{\psi_{i}}\left(\boldsymbol{r\times\gamma}\right)\psi_{i} \\
		&= Q_{i}\frac{R}{6}\frac{4\omega_{i}R+2m_{i}R-3}{2\omega_{i}R\left(\omega_{i}R-1\right)+m_{i}R},
	\end{aligned}\label{equ:mui}
\end{equation}
here, $Q_i$ represents the electric charge of the quark, $\boldsymbol{\gamma}$ is the Dirac matrix for spinor fields, and the parameters $(R,x_i)$
can be obtained by solving Eqs. (\ref{M}) and (\ref{transc}).
Furthermore, we present the expression for the hadron magnetic moment as follows:
\begin{equation}
	\mu =\left\langle \psi \left\vert \sum\nolimits_{i}2\mu
	_{i}S_{iz}\right\vert \psi \right\rangle ,  \label{musum}
\end{equation}
where $\psi$ is the color-spin wave function of the hadron.

The relevant parameters of the bag model are given in the Refs. \cite{Zhang:2021yul,Yan:2023lvm,Zhu:2023lbx} as 
\begin{equation}
	\begin{Bmatrix}
		Z_{0}=1.83,    & B^{1/4}=0.145\,\mathrm{GeV}, \\
		m_{n}=0\,\mathrm{GeV},    & m_{s}=0.279\,\mathrm{GeV}, \\
		m_{c}=1.641\,\mathrm{GeV},& m_{b}=5.093\,\mathrm{GeV}.
	\end{Bmatrix}
\end{equation} 

The final mass Eq. (\ref{M}) depends on the parameters $R$ and $x_i$, where $x_i$ represents the solution to the transcendental Eq. (\ref{transc}). Since $x_i$ itself depends on the variational parameter $R$, their values must be determined through iterative calculations.

\section{Singly Heavy pentaquarks in the Bag Model}\label{QCD limit}

 In Eq. (\ref{M}) contains three main elements of QCD: quark dynamics $\omega_i$, the perturbative part $H_{\text{CMI}} + H_{\text{EB}}$, and the non-perturbative QCD represented at the bag level. Most phenomenological studies include the first two components, such as the potential model, CMI model \cite{PhysRevD.12.147}, and naive quark model \cite{PhysRevD.17.3090}, while the effects of the QCD vacuum manifest in other parameters. Therefore, our current understanding of the non-perturbative contributions of hadrons and the confinement mechanism remains incomplete. However, the bag model can provide the confinement elements through the confinement energy, represented as 
\begin{equation}\label{VB}
	V_{B}= \frac{4}{3}\pi R^{3}B - \frac{Z_{0}}{R}.\\			
\end{equation}
The term $V_{B}$ is composed of two parts, where the parameter $B$ describes the difference between the disturbed and undisturbed QCD vacuum. This is analogous to how quarks interact in a hadron under the mean-field approximation within the hadronic bag, while the second term is contributed by the Casimir effect \cite{DeGrand:1975cf}. Together, these two components make up the confinement energy of the hadron. The bag parameters $B$ and $Z_{0}$ were determined in the original Ref. \cite{DeGrand:1975cf,Jaffe:1976ig} by fitting the nucleon's mass, charge radius, and magnetic moment, and were further validated in Ref. \cite{Zhang:2021yul} through additional fitting of other experimental results. Similar confinement images for the bag model are also reflected in Refs. \cite{PhysRevLett.119.091601,PhysRevLett.123.172301}. Here, we focus on the bag confinement term, extracting the second boundary conditions implicit in the bag model. We find that, although the color confinement provided by the bag model differs mathematically from the results obtained through lattice QCD fitting \cite{Bulava:2019iut,Bali:2005fu,Kou:2024dml}, their curves exhibit similar results at finite scales, as illustrated in Fig. \ref{fig:confinement}. The string picture in lattice QCD can describe the interactions between quarks, while the bag model presents a form of confinement based on hadronic degrees of freedom. For multiquark states, the complexity of color degrees of freedom increases the computational difficulty of lattice QCD; however, the bag model conveniently describes confinement features from a hadronic perspective. On the other hand, from the perspective of color interactions, there is a similarity between multiquark states and quark-antiquark pairs. In a multiquark configuration, we can always view the multiquark group as a single quark with color $3_c$
and another quark group with color $\bar{3}_c$ \cite{Liu:2025fbe}.

In Fig. \ref{fig:confinement}, we present the spin-independent bag radii of several hadrons: $R_B = 3.334 \, \text{GeV}^{-1}$, $R_D = 3.971 \, \text{GeV}^{-1}$, $R_{\Lambda_b/\Sigma_b} = 4.60 \, \text{GeV}^{-1}$, $R_{N/\Delta} = 5.252 \, \text{GeV}^{-1}$, and $R_{\Lambda_c/\Sigma_c} = 4.823\, \text{GeV}^{-1}$ \cite{Zhang:2021yul}. The bag radius for the singly heavy pentaquark system $nnnc\bar{n}$ is $R_{nnnc\bar{n}} = 6.00 \, \text{GeV}^{-1}$($1.18\, \text{fm}$). Lattice QCD indicates that the QCD string breaks within the range of $1.17\textendash1.29\, \text{fm}$, and similar conclusions from the bag model suggest that hadronic bags are unstable at a bag radius of $1.12\, \text{fm}$. As shown in the light red region of Fig. \ref{fig:confinement}, the bag radius for the $nnnc\bar{n}$ system lies precisely at the center of the ranges predicted by both the bag model and lattice estimates. Therefore, we estimate that under a compact picture, the $nnnc\bar{n}$ system will exhibit strong resonance characteristics due to the instability of the QCD vacuum, which is consistent with studies of mirror pentaquarks based on baryons \cite{PhysRevD.104.034009}.

\begin{figure}[h]
	\centering
	\includegraphics[width=0.45\textwidth]{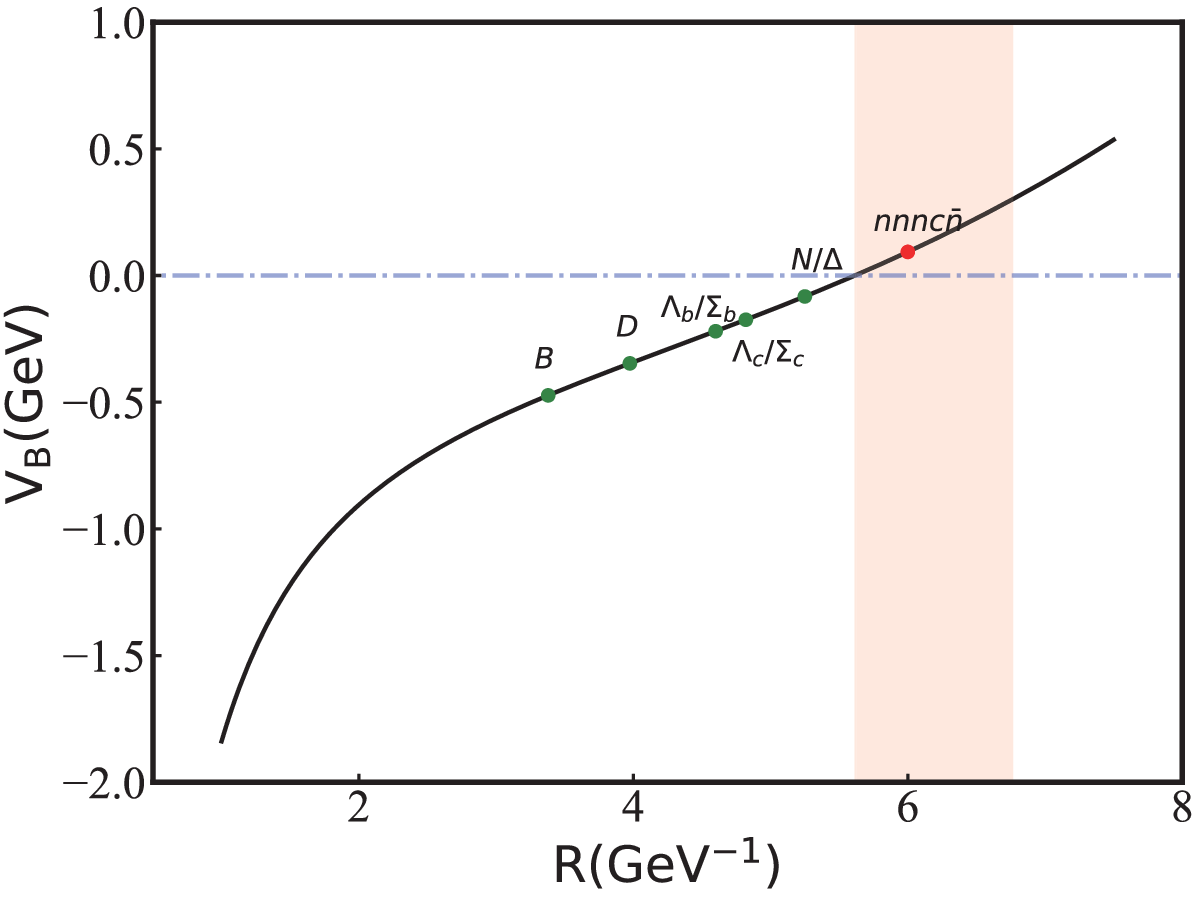}
	\caption{Bag confinement energy curve, with the red region corresponding to a resonance scale interval.}
	\label{fig:confinement}
\end{figure}

On the other hand, to distinguish from molecular states, we consider twice the bag radius of the $B$ meson, which corresponds to the contact radius of two $B$ mesons, as the lower limit for molecular states, approximately $6.67\, \text{GeV}^{-1}$($1.31\, \text{fm}$). This value nicely aligns with the upper limit for string breaking provided by lattice QCD. For the molecular configuration composed of $nnn$ and $c\bar{n}$, we estimate that its minimum radius should correspond to the contact radius of nucleons $N$ and $D$ mesons, which is about $9.22\, \text{GeV}^{-1}$ ($1.82\, \text{fm}$).

Here, we would like to clarify that our calculations for the $nnnQ\bar{n}$ system aim to explore its potential to exist in a resonant form. However, we cannot determine whether a specific $nnnQ\bar{n}$ system is a compact resonant state or a bound state with molecular characteristics. Based on our results, we believe that it is unlikely for multiquark hadrons with compact properties to exist in regions exceeding the complete string-breaking scale and above the bag confinement scale.

\section{Masses and Decays of Singly Heavy Pentaquarks}\label{sec:hadrons}

\subsection{Color-spin basis of pentaquarks}

For singly heavy pentaquarks, there are multiple flavor configurations. In our study, we focus solely on configurations that resemble the insertion of an $n\bar{n}$ pair into singly heavy baryons, forming mirror pentaquark structures. Specifically, these configurations include $nnnQ\bar{n}$ (where $n = u, d$ and $Q = c, b$), as well as those with strangeness $S = -1$ and $S = -2$, such as $snnQ\bar{n}$ and $ssnQ\bar{n}$.

In this context, we interpret the color structure of pentaquarks as the coupling of two substructures: one consisting of three quarks and the other being a quark-antiquark pair. Based on the group algebra of color symmetry $\textbf{SU(3)}_{c}$, we can derive three color wave functions. One of these is the color singlet \(\phi_{3}^{P}\), formed from the representation \(1 \otimes 1\), as shown in Eq. (\ref{8b3}), while the other two basis functions correspond to the color representation \(8 \otimes 8\), specifically \(\phi_{2}^{P}\) and \(\phi_{1}^{P}\), as presented in Eqs. (\ref{8b1}) and (\ref{8b2}).

For the spin, we continue to treat it as a combination of two substructures. The total spin is constructed using Clebsch-Gordan coefficients, and the spin basis vectors are denoted as $\chi_{i}$, as indicated in Eq. (\ref{Spin}) (where $i=1,2,3...10$). Finally, we present the color-spin basis vectors for the pentaquark state based on the group algebra of $\textbf{SU(3)}_{c} \otimes \textbf{SU(2)}_{s}$, as referenced in Eqs. (\ref{cs1})-(\ref{cs3}).

For the pentaquark system, there are various representations \cite{Zhang:2023hmg,Liu:2025fbe}. To facilitate the discussion of decays, we use final-state representations. For example, in the $nnsQ\bar{n}$ system, there are three final-state representations: $snQ \otimes n\bar{n}$, $nns \otimes Q\bar{n}$, and $nnQ \otimes s\bar{n}$. Due to the differences in symmetry, the choice of color-spin basis vectors also varies; however, the eigenvalues of mass and magnetic moment do not depend on the choice of basis vectors \cite{Liu:2025fbe}. When discussing decays, we reference different representations of the $Qsnn\bar{n}$ system; for instance, the representation $Qsn \otimes n\bar{n}$ corresponds to decay final states involving the $\Xi_{Q}$ family as well as $\pi/\omega/\rho$. Here, we consider all configurations of the substructures and provide a summary of the color-spin basis vectors for each substructure, organized in Table \ref{tab:basis}.

\subsection{S-wave two-body strong decay}

In the process of solving for the mass, we can obtain the eigenstates of our quark state under different substructure flavor configurations. Once the eigenstates of the pentaquark state are established, we can further discuss the two-body decay characteristics of the pentaquark state.

The form of the pentaquark state eigenstates obtained by solving the eigenvalue Eq. (\ref{M}) is as follows:
\begin{equation} \psi = c_{1}\phi_{1}^{P} \chi_{1}^{P} + c_{2}\phi_{2}^{P} \chi_{2}^{P} + c_{3}\phi_{3}^{P} \chi_{1}^{P} + \ldots. \label{equ:component} 
\end{equation}
This describes a color-spin mixed state, where the coefficients $|c_{i}|^2$ indicate the proportions of each color-spin configuration. The form of the eigenstate is related to the choice of basis vectors for the representation, as detailed in Table \ref{tab:basis}. In terms of decay, we only consider the proportion of the color-singlet state $\phi_{3}^{P}$, which determines the final decay state of the particle. For example, considering the pentaquarks with the flavor configuration $nnsQ\bar{n}$, where one representation is $nsQ \otimes n\bar{n}$, with spin-parity $J^P = 1/2^{-}$, the color-singlet state $\phi_{3}\chi_{9}$ specifies the decay final state as $\Xi_{c}\omega$. It is important to note that if the proportion of the color-singlet state $\phi_{3}^{P}$ satisfies $|c_{i}|^2 > 0.8$, this state is typically classified as a scattering state \cite{PhysRevD.105.074032,PhysRevD.103.034001}.

Next, we will discuss the two-body S-wave decay of the pentaquarks system. Drawing on these studies \cite{Weng:2019ynv,Weng:2020jao,gaoc1992,Weng:2021ngd,An:2020vku}, we will only consider S-wave two-body decays:
\begin{equation}
	\Gamma_{i}=\gamma_{i}\alpha\frac{k^{2L+1}}{m_{A}^{2L}}\cdot\left |c_{i}\right |^{2},
	\label{eq:eq12}
\end{equation} 
here, $\Gamma_{i}$ represents the decay width of channel $i$, while $\gamma_{i}$ is determined by the dynamics of the hadronic final state. $\alpha$ is the coupling constant. Since only S-wave decays are considered, the angular momentum term $L$ is set to zero. $m_{A}$  is the mass of the initial pentaquark state, $k$ is the momentum of the final state particles, and $|c_{i}|^{2}$ denotes the fraction of the eigenstate corresponding to the color representation $1 \otimes 1$. Here, we are only considering the $S$-wave decay of the pentaquark state, which is an OZI superallowed decay mode \cite{Jaffe:1976ig,Strottman:1979qu}, and the formula (\ref{eq:eq12}) simplifies to $\gamma_i\alpha k \cdot {|c_i|}^2$.

For the two-body decay process of a pentaquark state, such as $A \to B + C$, the momentum of the final state $k$ can be determined using the following equation:
\begin{equation}\label{phasespace}
	m_A = \sqrt{m_B^2 + k^2} + \sqrt{m_C^2 + k^2},
\end{equation}
where $m_B$ and $m_C$ represent the masses of the final state baryon and meson, respectively.

In Eq. (\ref{eq:eq12}), the parameter $\gamma_{i}$ depends on the spatial wave function of the final state. However, the quark model does not differentiate the influence of spin on the spatial wave function \cite{Weng:2019ynv}. Nonetheless, for the singly heavy pentaquark system, the effective coupling constants should be approximately consistent. In Refs. \cite{Zhang:2023hmg,Zhang:2023teh}, this characteristic width is close to the experimental observations, so I represent the decay width here as $k \cdot |c_i|^2$ \cite{Weng:2020jao}.

\subsection{$qqqQ\bar{q}$ Systems}

Next, we will discuss the mass spectrum of $nnnQ\bar{q}$. Here, we differentiate the mass spectrum based on the flavor symmetry of the substructure $nnn$, which is equivalent to distinguishing isospin. Specifically, $\{nnn\}$ denotes a flavor-symmetric configuration of the three quarks, while $[nnn]$ refers to a substructure that includes a flavor-antisymmetric light quark pair. According to heavy quark effective theory, the mass difference between the ground state baryons $\Sigma_Q$ and $\Lambda_Q$ in the heavy quark limit is equal to the difference arising from the light quark pair $nn$ (whether scalar or vector) \cite{Korner:1994nh}, reflecting the mass difference caused by the interactions between the flavor-symmetric $\{nn\}$ configuration and the flavor-antisymmetric $[nn]$ configuration \cite{Jia:2019bkr,PhysRevLett.119.202002,PhysRevD.92.074026}. Therefore, for the ground state baryons, we can obtain:
\begin{equation}\label{quark pair}
	\Sigma_{Q}(1/2^+) - \Lambda_{Q}(1/2^+) \approx E\{nn\} - E[nn] ,
\end{equation}
here, we take the average value from experimental data for $Q = b, c$ corresponding to the differences in ground state baryons, resulting in $E\{nn\} - E[nn] \approx 180 \text{ MeV}$.

For the pentaquarks, we consider the interactions among three light quarks. In the flavor-symmetric configuration of three quarks $\{nnn\}$, any interaction between two $nn$ pairs can be viewed as a flavor-symmetric $\{nn\}$, which means that at the interaction level, $\{nnn\}$ should satisfy $\{nnn\} \to 3\{nn\}$. Meanwhile, $[nnn]$ can be interpreted as containing one interaction of $\{nn\}$ and two interactions of $[nn]$, so we have $[nnn] \to \{nn\} + 2[nn]$. Therefore, we can simply extend the expression in Eq. (\ref{quark pair}) to the case of a singly pentaquark that includes interactions among three light quarks $n$ (where $n=u,d$), as follows:

\begin{equation}\label{tippair} 
	E\{nnn\} - E[nnn] = 2E\{nn\} - 2E[nn] \approx 360 \text{ MeV}. 
\end{equation}

Table \ref{1tab:nnnbn} presents the mass spectrum, magnetic moment, and bag radius of the $nnnb\bar{n}$ system. The mass range is between 6.2 and $6.9\, \text{GeV}$, with an average bag radius of $5.92\, \text{GeV}^{-1}$ ($1.17 \, \text{fm}$). We find that the mass of the $nnnb\bar{n}$ system is generally about $570\, \text{MeV}$ higher than the ground state mass of the baryon $\Lambda_b(5619)$, which is consistent with the predictions of chiral models \cite{Arifi:2020yfp,PhysRevD.104.034009}. Our calculated results are all above the corresponding lowest thresholds for meson-baryon combinations. For example, the minimum mass in our results, $P_{nnnb\bar{n}}(1/2^{-}, 6.195)$, is above the $\Lambda_b \pi$
threshold. In Table \ref{1tab:nnnbn}, we compare our results with various possible thresholds. Additionally, the bag radius of the pentaquark system shows a weak dependence on spin.

The pentaquarks exhibit more degrees of freedom compared to conventional baryons, with the ground state consisting of multiple states that share the same $J^P$
due to color-magnetic interactions. Therefore, we use the average value of the ground state to test the relation in Eq. (\ref{2tippair}). Here, the average energy for all $\{nnn\}$ configurations ($J^P=1/2^-$) is $\{\overline{P_{nnnb\bar{n}}}\} = 6793 \, \text{MeV}$, while the energy for the $[nnn]$ configuration ($J^P=1/2^-$) is $\overline{[P_{nnnb\bar{n}}]} = 6445 \, \text{MeV}$. This leads to the proposed ground state relation:
\begin{equation}\label{2tippair}
	\overline{\{P_{nnnQ\bar{n}}\}}-\overline{[P_{nnnQ\bar{n}}]} \approx	E\{nnn\} - E[nnn], 
\end{equation}		
the difference between $\overline{\{P_{nnnQ\bar{n}}\}}$ and $\overline{[P_{nnnb\bar{n}}]}$ equals $341 \, \text{MeV}$. Consequently, our findings are roughly consistent with the relationship specified by Eq. (\ref{tippair}).

\renewcommand{\tabcolsep}{0.1cm}
\renewcommand{\arraystretch}{0.7}
\begin{table*}[!htbp]
	\caption{The bag radius $R_{0}$ is measured in units of $\text{GeV}^{-1}$, the mass spectrum is expressed in GeV, and the magnetic moment is given in $\mu_N$, $\delta m$ is the mass calculated relative to the corresponding threshold energy.}
	\label{1tab:nnnbn}
	\begin{tabular}{cc|ccc|cccccccccc}
		\bottomrule[1.0pt]
		\multirow{2}{*}{F-symmetry} &\multirow{2}{*}{$J^{P}$} 
		&\multicolumn{3}{l|}{$P_{nnnb\bar{n}}$} 
		&\multicolumn{10}{l}{$\delta m=M-M_{Threshold}$} \\
		& &$R_{0}$ &$M$ &$\mu$ 
		&$\Sigma_{b}^{\ast}\omega$ &$\Sigma_{b}\omega$ &$\Sigma_{b}^{\ast}\pi$ &$\Sigma_{b}\pi$ &$\Lambda_{b}\omega$ &$\Lambda_{b}\pi$ 
		&$\Delta B^{\ast}$ &$\Delta B$ &$NB^{\ast}$ &$NB$ \\ \hline
		$\{nnn\}$
		&${5/2}^{-}$    &5.94   &6.735  &7.33, 4,15, 0.97, -2.20, -5.38 &0.119 & & & & & &0.178 & & & \\
		&${3/2}^{-}$    &5.95   &6.852  &3.93, 1.33, 0.74, 0.15, -2.45 &0.236 &0.255 &0.879 & & & &0.295 &0.340 &0.588 & \\
		&               &5.91   &6.714  &6.55, 0.78, 0.92, 1.06, -4.71 &0.098 &0.117 &0.741 & & & &0.157 &0.202 &0.450 & \\
		&               &5.88   &6.571  &3.96, 1.36, 0.76, 0.15, -2.45 &-0.045 &-0.026 &0.598 & & & &0.014 &0.059 &0.307 & \\
		&${1/2}^{-}$    &5.99   &6.954  &0.15, 0.39, 0.08, -0.22, 0.01 &0.338 &0.357 & &1.000 & &1.194 &0.397 & &0.690 &0.735 \\
		&               &5.96   &6.860  &2.53, 0.68, 0.48, 0.26, -1.58 &0.244 &0.263 & &0.906 & &1.100 &0.303 & &0.596 &0.641 \\
		&               &5.87   &6.563  &2.56, 0.66, 0.47, 0.28, -1.62 &-0.053 &-0.034 & &0.609 & &0.803 &0.006 & &0.299 &0.344 \\
		\hline   
		$[nnn]$
		&${5/2}^{-}$    &5.94   &6.735  &4.15, 0.97, -2.20 &0.119 & & & & & &0.178 & & & \\
		&${3/2}^{-}$    &5.93   &6.726  &2.62, 0.88, -0.86 &0.110 &0.129 &0.753 & &0.323 & &0.169 &0.214 &0.462 & \\
		&               &5.90   &6.705  &3.25, 0.60, -2.06 &0.089 &0.108 &0.732 & &0.302 & &0.148 &0.193 &0.441 & \\
		&               &5.87   &6.489  &2.01, 0.09, -1.83 &-0.127 &-0.108 &0.516 & &0.086 & &-0.068 &-0.023 &0.225 & \\
		&               &5.81   &6.236  &1.99, 0.75, -0.50 &-0.380 &-0.361 &0.263 & &-0.167 & &-0.321 &-0.276 &-0.028 & \\
		&${1/2}^{-}$    &5.91   &6.709  &1.42, 0.44, -0.54 &0.093 &0.112 & &0.755 &0.306 &0.949 &0.152 & &0.445 &0.490 \\
		&               &5.88   &6.574  &-0.06, -0.19, -0.32 &-0.042 &-0.023 & &0.620 &0.171 &0.814 &0.017 & &0.310 &0.355 \\
		&               &5.85   &6.476  &1.41, 0.15, -1.12 &-0.140 &-0.121 & &0.522 &0.073 &0.716 &-0.081 & &0.212 &0.257 \\
		&               &5.80   &6.274  &-0.05, 0.00, 0.05 &-0.342 &-0.323 & &0.320 &-0.129 &0.514 &-0.283 & &0.010 &0.055 \\
		&               &5.74   &6.195  &1.36, 0.56, -0.25 &-0.421 &-0.402 & &0.241 &-0.208 &0.435 &-0.362 & &-0.069 &-0.024 \\
		\bottomrule[1.0pt]
	\end{tabular}
\end{table*}

\renewcommand{\tabcolsep}{0.32cm}
\renewcommand{\arraystretch}{0.8}
\begin{table*}[!htbp]
	\caption{The characteristic width $|c_i|^2 \cdot k$ is measured in units of GeV, $\ast$ denotes scattering states, and $\cdots$ indicates threshold suppression.}
	\label{2tab:nnnbn}
	\begin{tabular}{cc|cccccc|cccc}
		\bottomrule[1.0pt]
		\multicolumn{2}{c|}{} & \multicolumn{6}{c|}{$nnb\otimes n\bar{n}$} & \multicolumn{4}{c}{$nnn\otimes b\bar{n}$} \\
		$J^{P}$ & $M$ & $\Sigma_{b}^{\ast}\omega/\rho$ & $\Sigma_{b}\omega/\rho$ & $\Sigma_{b}^{\ast}\pi$ & $\Sigma_{b}\pi$ & $\Lambda_{b}\omega/\rho$ & $\Lambda_{b}\pi$ & $\Delta B^{\ast}$ & $\Delta B$ & $NB^{\ast}$ & $NB$ \\
		\hline
		${5/2}^{-}$ & 6.735 & $\ast$ & & & & & & & & & \\
		${3/2}^{-}$ & 6.852 & 0.130 & 0.030 & 0.004 & & & & & 0.152 & & \\
		& 6.714 & 0.013 & 0.032 & 0.005 & & & & & 0.346 & & \\
		& 6.571 & $\cdots$ & $\cdots$ & 0.145 & & & & & 0.101 & & \\
		${1/2}^{-}$ & 6.954 & 0.255 & 0.069 & & $<10^{-4}$ & & & 0.017 & & & \\
		& 6.860 & 0.021 & 0.143 & & 0.002 & & & 0.449 & & & \\
		& 6.563 & $\cdots$ & $\cdots$ & & 0.211 & & & 0.046 & & & \\
		\hline
		${5/2}^{-}$ & 6.735 & & & & & & &$\ast$ & & & \\
		${3/2}^{-}$ & 6.726 & & & & & 0.169 & & & & 0.003 & \\
		& 6.705 & & & & & 0.095 & & & & $<10^{-4}$ & \\
		& 6.489 & & & & & 0.211 & & & & 0.240 & \\
		& 6.236 & & & & & 0.016 & & & & $\cdots$ & \\
		${1/2}^{-}$ & 6.709 & & & & & 0.199 & 0.001 & & & & 0.002 \\
		& 6.574 & & & & & 0.001 & 0.047 & & & & 0.037 \\
		& 6.476 & & & & & 0.079 & 0.003 & & & & 0.184 \\
		& 6.274 & & & & & $\cdots$ & 0.295 & & & & 0.017 \\
		& 6.195 & & & & & $\cdots$ & 0.005 & & & & $\cdots$ \\
		\bottomrule[1.0pt]
	\end{tabular}
\end{table*}

In Table \ref{2tab:nnnbn}, we analyze the strong decay stability of the $nnnb\bar{n}$ systems. Here, we study these systems based on the $nnQ \otimes n\bar{n}$ and $nnn \otimes Q\bar{n}$ representations, which encompass all S-wave strong decay channels of the $nnnb\bar{n}$ system. We describe the characteristics of the decay widths using the expression $k \cdot |c_i|^2$ from Eq. (\ref{eq:eq12}). Additionally, under similar conditions \cite{Weng:2019ynv,Weng:2021ngd,An:2020vku}, the effects of the spins of the final state hadrons on the spatial wave function can be neglected, meaning that the $\gamma_{i}$
in Eq. (\ref{eq:eq12}) is independent of the spins of the final state hadrons. For the $nnnb\bar{n}$ system, the following similarity relation holds:
\begin{equation}\label{similar}
	\begin{aligned}
		&\gamma_{\Sigma_{b}^{\ast}\omega/\rho}=\gamma_{\Sigma_{b}\omega/\rho}=\gamma_{\Sigma_{b}^{\ast}\pi}=\gamma_{\Sigma_{b}\pi}\\
		&\gamma_{\Lambda_{b}\omega/\rho} = \gamma_{\Lambda_{b}\pi}, \quad \gamma_{\Delta B^{\ast}} = \gamma_{\Delta B^{\ast}}, \\
		&\gamma_{NB^{\ast}} = \gamma_{NB}. \\
	\end{aligned}
\end{equation}

Based on the above relations and Eq. (\ref{eq:eq12}), we can analyze the decay width ratios for the decay channels. For example, for $P_{nnnb\bar{n}}(3/2^-, 6.852)$, the decay final states of $\Sigma_{b}^{\ast}\pi$, $\Sigma_{b}\omega$, and $\Sigma_{b}^{\ast}\rho$
can have their decay width ratios expressed in terms of the values of $k \cdot |c_i|^2$, specifically given as $(0.130:0.030:0.004) = (1:0.23:0.03)$. Next, we will represent the characteristic partial widths using the value of $k \cdot |c_i|^2$ \cite{Weng:2020jao}.
We have observed that, influenced by threshold effects, certain states exhibit decay modes limited to a single channel, such as $P_{nnnb\bar{n}}(3/2^-, 6.236)$ and $P_{nnnb\bar{n}}(1/2^-, 6.195)$, both of which have relatively small values of $k \cdot |c_i|^2$, on the order of tens of MeV. Furthermore, the masses of these states are elevated by approximately $500\textendash600\, \text{MeV}$ above the ground state $\Lambda_b (5619)$, aligning with the fundamental criteria for pentaquarks formation and potentially piquing experimental interest due to their relative stability. In contrast, the majority of other $P_{nnnb\bar{n}}$ states exhibit multiple strong decay channels, with characteristic partial widths exceeding $150\, \text{MeV}$. Consequently, compared to other excited states of $\Lambda_b$, the majority of $P_{nnnb\bar{n}}$ states demonstrate significant strong decay instability.

\renewcommand{\tabcolsep}{0.1cm}
\renewcommand{\arraystretch}{0.7}
\begin{table*}[!htbp]
	\caption{The bag radius $R_{0}$ is measured in units of $\text{GeV}^{-1}$, the mass spectrum is expressed in GeV, and the magnetic moment is given in $\mu_N$, $\delta m$ is the mass calculated relative to the corresponding threshold energy.}
	\label{tab1:nnncn}
	\begin{tabular}{cc|ccc|cccccccccc}
		\bottomrule[1.0pt]
		\multirow{2}{*}{F-symmetry} &\multirow{2}{*}{$J^{P}$} 
		&\multicolumn{3}{l|}{$P_{nnnc\bar{n}}$} 
		&\multicolumn{10}{l}{$\delta m=M-M_{Threshold}$} \\
		& &$R_{0}$ &$M$ &$\mu$ 
		&$\Sigma_{c}^{\ast}\omega$ &$\Sigma_{c}\omega$ &$\Sigma_{c}^{\ast}\pi$ &$\Sigma_{c}\pi$ &$\Lambda_{c}\omega$ &$\Lambda_{c}\pi$ 
		&$\Delta D^{\ast}$ &$\Delta D$ &$ND^{\ast}$ &$ND$ \\ \hline
		$\{nnn\}$
		&${5/2}^{-}$    &6.07   &3.348  &8.09, 4.84, 1.59, -1.66, -4.92 &0.047 & & & & & &0.107 & & & \\
		&${3/2}^{-}$    &6.06   &3.442  &4.55, 1.79, 1.29, 0.80, -1.96 &0.141 &0.205 &0.784 & & & &0.201 &0.342 &0.494 & \\
		&               &6.05   &3.310  &5.90, 0.17, 0.55, 0.93, -4.80 &0.009 &0.073 &0.652 & & & &0.069 &0.210 &0.362 & \\
		&               &5.94   &3.147  &5.09, 2.39, 1.44, 0.49, -2.21 &-0.154 &-0.090 &0.489 & & & &-0.094 &0.047 &0.199 & \\
		&${1/2}^{-}$    &6.14   &3.569  &1.24, 1.57, 0.83, 0.07, 0.40 &0.268 &0.332 & &0.975 & &1.143 &0.328 & &0.621 &0.762 \\
		&               &6.05   &3.451  &1.95, 0.12, 0.21, 0.30, -1.53 &0.150 &0.214 & &0.857 & &1.025 &0.210 & &0.503 &0.644 \\
		&               &5.97   &3.144  &2.34, 0.27, 0.22, 0.16, -1.92 &-0.157 &-0.093 & &0.550 & &0.718 &-0.097 & &0.196 &0.337 \\
		\hline 
		$[nnn]$
		&${5/2}^{-}$    &6.07   &3.348  &4.84, 1.59, -1.66 &0.047 & & & & & &0.107 & & & \\
		&${3/2}^{-}$    &6.07   &3.338  &2.88, 1.25, -0.39 &0.037 &0.101 &0.680 & &0.269 & &0.097 &0.238 &0.390 & \\
		&               &5.98   &3.270  &3.32, 0.54, -2.24 &-0.031 &0.033 &0.612 & &0.201 & &0.029 &0.170 &0.322 & \\
		&               &5.99   &3.092  &2.66, 1.27, -1.39 &-0.209 &-0.145 &0.434 & &0.023 & &-0.149 &-0.008 &0.144 & \\
		&               &5.96   &2.957  &2.63, 1.38, 0.11 &-0.444 &-0.380 &0.199 & &-0.212 & &-0.384 &-0.243 &-0.091 & \\
		&${1/2}^{-}$    &6.04   &3.304  &1.20, 0.37, -0.47 &0.003 &0.067 & &0.710 &0.235 &0.878 &0.063 & &0.356 &0.497 \\
		&               &5.97   &3.160  &0.58, 0.20, -0.19 &-0.141 &-0.077 & &0.566 &0.091 &0.734 &-0.081 & &0.212 &0.353 \\
		&               &5.96   &3.060  &1.21, -0.08, -1.38 &-0.241 &-0.177 & &0.466 &-0.009 &0.634 &-0.181 & &0.112 &0.253 \\
		&               &5.92   &2.871  &0.58, 0.69, 0.79 &-0.430 &-0.366 & &0.277 &-0.198 &0.445 &-0.370 & &-0.077 &0.064 \\
		&               &5.79   &2.738  &1.16, 0.39, -0.38 &-0.563 &-0.499 & &0.144 &-0.331 &0.312 &-0.503 & &-0.210 &-0.069 \\
		\bottomrule[1.0pt]
	\end{tabular}
\end{table*}

\renewcommand{\tabcolsep}{0.32cm}
\renewcommand{\arraystretch}{0.8}
\begin{table*}[!htbp]
	\caption{The characteristic width $|c_i|^2 \cdot k$ is measured in units of GeV, $\ast$ denotes scattering states, and $\cdots$ indicates threshold suppression.}
	\label{tab2:nnncn}
	\begin{tabular}{cc|cccccc|cccc}
		\bottomrule[1.0pt]
		\multicolumn{2}{c|}{} & \multicolumn{6}{c|}{$nnc\otimes n\bar{n}$} & \multicolumn{4}{c}{$nnn\otimes c\bar{n}$} \\
		 $J^{P}$ &$M$ 
		&$\Sigma_{c}^{\ast}\omega/\rho$ &$\Sigma_{c}\omega/\rho$ &$\Sigma_{c}^{\ast}\pi$ &$\Sigma_{c}\pi$ &$\Lambda_{c}\omega/\rho$ &$\Lambda_{c}\pi$ 
		&$\Delta D^{\ast}$ &$\Delta D$ &$ND^{\ast}$ &$ND$ \\ \hline
	
		${5/2}^{-}$      &3.348 &$\ast$ & & & & & & & & & \\
		${3/2}^{-}$      &3.442 &0.089&0.028&0.004& & & & &0.106 & & \\
		                 &3.310 &0.007&0.024&0.029& & & & &0.181 & & \\
		                 &3.147 &$\cdots$&$\cdots$&0.097& & & & &0.144 & & \\
		${1/2}^{-}$      &3.569 &0.216&0.020& &$<10^{-4}$& & &0.084 & & & \\
		                 &3.451 &$<10^{-4}$&0.144& &$<10^{-4}$& & &0.293 & & & \\
		                 &3.144 &$\cdots$&$\cdots$& &0.133& & &$\cdots$ & & & \\
		\hline   
		${5/2}^{-}$     &3.348 & & & & & & &$\ast$ & & & \\
		${3/2}^{-}$     &3.338 & & & & &0.099& & & &0.188 & \\
		                &3.270 & & & & &0.046& & & &$<10^{-4}$& \\
		                &3.092 & & & & &0.043& & & & 0.016& \\
		                &2.856 & & & & &$\cdots$& & & &$\cdots$& \\
		${1/2}^{-}$     &3.304 & & & & &0.242&0.003& & & &0.003 \\
		                &3.160 & & & & &0.005&0.034& & & &0.217 \\
		                &3.060 & & & & &$\cdots$&0.019& & & &0.138 \\
		                &2.871 & & & & &$\cdots$&0.229 & & & &0.019\\
		                &2.738 & & & & &$\cdots$&0.011& & & &$\cdots$\\
		\bottomrule[1.0pt]
	\end{tabular}
\end{table*}

In Table \ref{tab1:nnncn}, we present the mass spectra, magnetic moments, and bag radii for $nnnc\bar{n}$. The average bag radiu of the hadrons is approximately $6.00\, \text{GeV}^-1$($1.18 \, \text{fm}$), with their masses ranging from $2.7\textendash3.6 \, \text{GeV}$. For the $nnnc\bar{n}$ system, the lowest mass state $P_{nnnc\bar{n}}(1/2^{-}, 2738)$ is found to be approximately $452 \, \text{MeV}$ above the $\Lambda_c (2286)$. A negative parity state $\Lambda_c (1/2^-, 3529)$ has been predicted in the Ref. \cite{PhysRevD.104.034009} and confirmed to be dominated by pentaquarks configurations. Our results correspond nicely with the quantum numbers of $P_{nnnc\bar{n}}(1/2^{-}, 3569)$.

Similar to the $nnnb\bar{n}$ system, for the flavor-symmetric $\{nnn\}$ state with $J^P=1/2^-$, the average value is $\{\overline{P_{nnnc\bar{n}}}\} = 3388\, \text{MeV}$. In the case of the flavor-antysymmetric configuration $\{nnn\}$, the average value is $[\overline{P_{nnnc\bar{n}}}] = 3027\, \text{MeV}$, resulting in a difference of $361\, \text{MeV}$, which precisely satisfies the relationship described in Eq. (\ref{tippair}).

In Table \ref{tab2:nnncn}, we analyze the strong decay widths of the $nnnc\bar{n}$ system, primarily based on the values of $k \cdot |c_i|^2$. Additionally, for strong decay channels that satisfy similarity relations, such as those in the $nnnb\bar{n}$ system as described in Eq. (\ref{similar}), the decay width ratios can be described by the ratios of $k \cdot |c_i|^2$. The $nnnc\bar{n}$ system is generally unstable with respect to strong decays, with most states having maximum decay widths exceeding $100, \text{MeV}$. However, some states exhibit relatively fewer decay modes and narrower widths due to color and threshold effects, such as $P_{nnnc\bar{n}}(1/2^-, 3144)$,\\   $P_{nnnc\bar{n}}(1/2^-, 3060)$, and $P_{nnnc\bar{n}}(1/2^-, 2738)$. These states are more likely to be observed experimentally compared to others. Additionally, we find that $P_{nnnc\bar{n}}(5/2^-, 6.852)$corresponds to a scattering state formed by $\Sigma_{c}^{\ast}\omega/\rho$.

\subsection{$qqsQ\bar{q}$ Systems}

Next, we consider the $nnsQ\bar{n}$
system with strangeness $S = -1$, where the flavor structure adheres to $SU(3)_{F}$ symmetry. We distinguish the mass spectrum based on the symmetric and antisymmetric configurations of the $nns$ substructure. Specifically, we examine two scenarios: one involves the flavor-symmetric configuration $\{snn\}$, while the other includes the flavor-antisymmetric case $[snn]$. Similar to Eq. (\ref{quark pair}), for the ground state baryon with strangeness $s = -1$ and $J^p = 1/2^+$, the following approximate relation holds:
\begin{equation}\label{squark pair} 
	\Xi_{Q}^{\prime} - \Xi_{Q} \approx E\{ns\} - E[sn] \approx 120\text{ MeV},
\end{equation}
by comparing with Eq. (\ref{tippair}), we can map the above relationship to the pentaquark system with strangeness $s = -1$:
\begin{equation}\label{s2tippair} 
	\{\overline{P_{nnsQ\bar{n}}}\} - \overline{[P_{nnsQ\bar{n}}]} \approx E\{snn\} - E[nns]\approx240\text{ MeV}.
\end{equation}

In Table \ref{1tab:nnsbn}, we present the mass spectrum of $nnsb\bar{n}$. The weakening of flavor symmetry leads to multiple states corresponding to this system. The average bag radiu is approximately $5.9 \, \text{GeV}^{-1}$($1.16 \, \text{fm}$), which corresponds to a strong resonance scale. The mass range is roughly between 6.1 and $7.0\, \text{GeV}$, with most states having a mass more than 500 MeV above the ground state baryon $\Xi_{b}(5791)$, and all states lying at or above the threshold corresponding to baryon-meson combinations. Additionally, the difference between the average values $\{\overline{P_{nnsb\bar{n}}}\}$ and $\overline{[P_{nnsb\bar{n}}]}$ for $J^P = 1/2^-$ in our calculations is $210 \, \text{MeV}$, which is approximately equal to the value given by Eq. (\ref{s2tippair}).

In Table \ref{2tab:nnsbn}, we analyze the strong decay stability of the $nnsb\bar{n}$ system. Overall, the characteristic widths of the main decay channels for the $nnsb\bar{n}$ system exceed $120\, \text{MeV}$. However, it appears that the strong decay stability of the $nnsb\bar{n}$ system is greater than that of the $nnnb\bar{n}$ system, as reflected in the partial decay widths. This is because the flavor symmetry of $nnsb\bar{n}$ is lower than that of $nnnb\bar{n}$; as flavor symmetry decreases, the characteristic widths for strong decays tend to diminish overall. The results reveal two scattering states: $P_{nnsb\bar{n}}(5/2^{-}, 6889)$ and $P_{nnsb\bar{n}}(5/2^{-}, 6857)$, with final scattering states of $\Sigma_{b}K^{\ast}$ and $\Xi_{b}^{\ast}\omega$, respectively. Additionally, certain states that are stable against strong decay are also noteworthy, such as  $P_{nnsb\bar{n}}(1/2^{-}, 6107)$, $P_{nnsb\bar{n}}(1/2^{-}, 6621)$,\\ and $P_{nnsb\bar{n}}(3/2^{-}, 6360)$.

\renewcommand{\tabcolsep}{0.11cm}
\renewcommand{\arraystretch}{0.7}
\begin{table*}[!htbp]
	\caption{The bag radius $R_{0}$ is measured in units of $\text{GeV}^{-1}$, the mass spectrum is expressed in GeV, and the magnetic moment is given in $\mu_N$, $\delta m$ is the mass calculated relative to the corresponding threshold energy.}
	\label{1tab:nnsbn}
	\begin{tabular}{cc|cc|cccccccccccc}
		\bottomrule[1.0pt]
		\multirow{2}{*}{F-symmetry} &\multirow{2}{*}{$J^{P}$} 
		&\multicolumn{2}{l|}{$P_{nnsb\bar{n}}\quad R_{0}=5.90\,$GeV$^{-1}$}
		&\multicolumn{12}{l}{$\delta m=M-M_{Threshold}$} \\
		& &$M$ &$\mu$ &$\Xi_{b}^{\ast}\omega$ &$\Xi_{b}\omega$ &$\Xi_{b}^{\ast}\pi$ &$\Xi_{b}\pi$ 
		&$\Sigma_{b}^{\ast}K^{\ast}$ &$\Sigma_{b}K^{\ast}$ &$\Sigma_{b}^{\ast}K$ &$\Sigma_{b}K$  
		&$\Lambda_{b}K^{\ast}$ &$\Lambda_{b}K$ &$\Lambda B^{\ast}$ &$\Lambda B$ \\ \hline
		$\{nns\}$
		&${5/2}^{-}$    &6.889  &4.40, 1.25, -1.91, -5.07 &0.152 & & & &0.162 & & & & & & & \\
		&               &6.857  &4.40, 1.25, -1.91, -5.07 &0.120 & & & &0.130 & & & & & & & \\
		&${3/2}^{-}$    &6.975  &2.09, 0.91, -0.50, -2.17 &0.238 &0.398 &0.881 & &0.248 &0.267 &0.646 & &0.461 & &0.534 & \\
		&               &6.876  &2.15, 0.65, -1.17, -2.67 &0.139 &0.299 &0.782 & &0.149 &0.168 &0.547 & &0.362 & &0.435 & \\
		&               &6.865  &3.49, 0.81, -1.58, -4.25 &0.128 &0.288 &0.771 & &0.138 &0.157 &0.536 & &0.351 & &0.424 & \\
		&               &6.834  &3.56, 0.81, -1.63, -4.38 &0.097 &0.257 &0.740 & &0.107 &0.126 &0.505 & &0.320 & &0.393 & \\
		&               &6.718  &2.95, 1.34, -0.87, -2.48 &-0.019 &0.141 &0.624 & &-0.009 &0.010 &0.389 & &0.204 & &0.277 & \\
		&               &6.680  &4.02, 1.51, -0.70, -3.21 &-0.057 &0.103 &0.586 & &-0.047 &-0.028 &0.351 & &0.166 & &0.239 & \\
		&               &6.453  &2.68, 1.36, -0.88, -2.20 &-0.284 &-0.124 &0.359 & &-0.274 &-0.255 &0.124 & &-0.061 & &0.012 & \\
		&${1/2}^{-}$    &7.055  &0.06, 0.18, -0.09, 0.03 &0.318 &0.478 & &1.121 &0.328 &0.347 & &0.745 &0.541 &0.939 &0.614 &0.659 \\
		&               &6.983  &1.33, 0.49, -0.57, -1.40 &0.246 &0.406 & &1.049 &0.256 &0.275 & &0.673 &0.469 &0.867 &0.542 &0.587 \\
		&               &6.862  &0.91, 0.13, -0.67, -1.45 &0.125 &0.285 & &0.928 &0.135 &0.154 & &0.552 &0.348 &0.746 &0.421 &0.466 \\
		&               &6.768  &-0.31, -0.41, -0.16, -0.26 &0.031 &0.191 & &0.834 &0.041 &0.060 & &0.458 &0.254 &0.652 &0.327 &0.372 \\
		&               &6.709  &2.00, 0.85, -0.49, -1.65 &-0.028 &0.132 & &0.775 &-0.018 &0.001 & &0.399 &0.195 &0.593 &0.268 &0.313 \\
		&               &6.665  &2.82, 1.16, -0.37, -2.02 &-0.072 &0.088 & &0.731 &-0.062 &-0.043 & &0.355 &0.151 &0.549 &0.224 &0.269 \\
		&               &6.444  &-0.55, -0.37, -0.24, -0.05 &-0.293 &-0.133 & &0.510 &-0.283 &-0.264 & &0.134 &-0.070 &0.328 &0.003 &0.048 \\
		&               &6.416  &2.42, 1.39, -0.33, -1.37 &-0.321 &-0.161 & &0.482 &-0.311 &-0.292 & &0.106 &-0.098 &0.300 &-0.025 &0.020 \\
		\hline
		$[nns]$
		&${5/2}^{-}$    &6.857  &1.91, -1.25 &0.120 & & & &0.130 & & & & & & & \\
		&${3/2}^{-}$    &6.843  &0.66, -0.87 &0.106 &0.266 &0.749 & &0.116 &0.135 &0.514 & &0.329 & &0.402 & \\
		&               &6.827  &1.68, -1.18 &0.090 &0.250 &0.733 & &0.100 &0.119 &0.498 & &0.313 & &0.386 & \\
		&               &6.653  &0.13, -2.61 &-0.084 &0.076 &0.559 & &-0.074 &-0.055 &0.324 & &0.139 & &0.212 & \\
		&               &6.621  &0.96, -2.14 &-0.116 &0.044 &0.527 & &-0.106 &-0.087 &0.292 & &0.107 & &0.180 & \\
		&               &6.360  &-0.24, -0.76 &-0.377 &-0.217 &0.266 & &-0.367 &-0.348 &0.031 & &-0.154 & &-0.081 & \\
		&${1/2}^{-}$    &6.824  &0.71, -0.19 &0.087 &0.247 & &0.890 &0.097 &0.116 & &0.514 &0.310 &0.708 &0.383 &0.428 \\
		&               &6.705  &0.16, -0.26 &-0.032 &0.128 & &0.771 &-0.022 &-0.003 & &0.395 &0.191 &0.589 &0.264 &0.309 \\
		&               &6.637  &0.16, -1.62 &-0.100 &0.060 & &0.703 &-0.090 &-0.071 & &0.327 &0.123 &0.521 &0.196 &0.241 \\
		&               &6.621  &0.69, -1.33 &-0.116 &0.044 & &0.687 &-0.106 &-0.087 & &0.311 &0.107 &0.505 &0.180 &0.225 \\
		&               &6.475  &-0.15, 0.10 &-0.262 &-0.102 & &0.541 &-0.252 &-0.233 & &0.165 &-0.039 &0.359 &0.034 &0.079 \\
		&               &6.325  &-0.10, -0.47 &-0.412 &-0.252 & &0.391 &-0.402 &-0.383 & &0.015 &-0.189 &0.209 &-0.116 &-0.071 \\
		&               &6.107  &-0.09, -0.12 &-0.630 &-0.470 & &0.173 &-0.620 &-0.601 & &-0.203 &-0.407 &-0.009 &-0.334 &-0.289 \\
		\bottomrule[1.0pt]
	\end{tabular}
\end{table*}

\renewcommand{\tabcolsep}{0.05cm}
\renewcommand{\arraystretch}{0.8}
\begin{table*}[!htbp]
	\caption{The characteristic width $|c_i|^2 \cdot k$ is measured in units of GeV, $\ast$ denotes scattering states, and $\cdots$ indicates threshold suppression.}
	\label{2tab:nnsbn}
	\begin{tabular}{cc|cccccc|cccccc|cccccc}
		\bottomrule[1.0pt]
			\multicolumn{2}{c|}{} & \multicolumn{6}{c|}{$nsb\otimes n\bar{n}$} &\multicolumn{6}{c|}{$nnb\otimes s\bar{n}$} & \multicolumn{6}{c}{$nns\otimes b\bar{n}$} \\
		$J^{P}$&$M$ &$\Xi_{b}^{\ast}\omega/\rho$ &$\Xi_{b}\omega/\rho$&$\Xi_{b}^{\prime}\omega/\rho$ &$\Xi_{b}^{\prime}\pi$&$\Xi_{b}^{\ast}\pi$ &$\Xi_{b}\pi$ 
		&$\Sigma_{b}^{\ast}K^{\ast}$ &$\Sigma_{b}K^{\ast}$ &$\Sigma_{b}^{\ast}K$ &$\Sigma_{b}K$  
		&$\Lambda_{b}K^{\ast}$ &$\Lambda_{b}K$ &$\Lambda B^{\ast}$ &$\Lambda B$&$\Sigma B^{\ast}$&$\Sigma B$&$\Sigma^{\ast} B$&$\Sigma^{\ast} B^{\ast}$ \\ \hline
		${5/2}^{-}$    &6.889 & & & & & & &$\ast$ & & & & & & & & & & & \\
		&6.857 &$\ast$ & & & & & & & & & & & & & & & & &  \\
		${3/2}^{-}$    &6.975 &0.107 &$<10^{-4}$ &0.028 & &    0.002   & &0.159 & 0.037&$<10^{-4}$& & & & & &0.002 & &0.156 &0.318  \\
		&6.876 &0.075 &0.096 &0.001 & &0.001 &  &0.311 &0.008 &0.018 && &  & & &0.012 & &0.017 &0.008  \\
		&6.865 &0.022&0.016 &$<10^{-4}$ & &0.002 & &0.004 &0.233 &0.013 & & & & & &$<10^{-4}$ & & 0.265&0.183  \\
		&6.834 & 0.010 & 0.011 &0.123 & &0.001 & &0.001 &0.192 &$<10^{-4}$ & & & & & &$<10^{-4}$ & &0.088 &0.054  \\
		&6.718 &$\cdots$& 0.006&$\cdots$& &0.100 & &$\cdots$&0.003 &0.250 & & & & & &$<10^{-4}$ & &0.086 &0.023  \\
		&6.680 &$\cdots$&0.076 &$\cdots$& &0.010 & &$\cdots$&$\cdots$&0.007 & & & & & &0.470 & &0.001 &$\cdots$ \\
		&6.453 &$\cdots$&$\cdots$&$\cdots$& &0.116 & &$\cdots$&$\cdots$ &0.207 & & & & & &0.470 & &$\cdots$&$\cdots$  \\
		${1/2}^{-}$    &7.055 &0.244 &$<10^{-4}$ &0.059 &$<10^{-4}$ & & $<10^{-4}$& 0.284&0.063 & & $<10^{-4}$& & & & &$<10^{-4}$ &$<10^{-4}$ & &0.027\\
		&6.983 &0.013&$<10^{-4}$&0.135 &$<10^{-4}$ & &$<10^{-4}$ & 0.017&0.171 & &$<10^{-4}$ & & & & &0.001 &0.001 & &0.473  \\
		&6.862&0.035 &0.135 &0.044 &$<10^{-4}$ & &$<10^{-4}$ &0.114 &0.147 & &0.020 & & & & &0.003 &0.010 & &0.001 \\
		&6.768 &0.012 &0.001 &0.019 &$<10^{-4}$ & &0.018 &0.056 &0.067& &0.001 & & & & &0.352 &0.078 &  &$<10^{-4}$ \\
		&6.709 &$\cdots$&0.005 &$\cdots$&0.109 & &$<10^{-4}$ &$\cdots$ &0.003 & &0.252 & & & & &$<10^{-4}$ &$<10^{-4}$ & &$\cdots$\\
		&6.665 &$\cdots$ &0.064 &$\cdots$&0.014 & &0.002 &$\cdots$ &$\cdots$ & &0.013 & & & & &0.103 &0.377 & &$\cdots$\\
		&6.444 &$\cdots$ &$\cdots$&$\cdots$ &0.013 & &0.214 &$\cdots$&$\cdots$& &0.022 & & & & &$\cdots$&$\cdots$& &$\cdots$ \\
		&6.416 &$\cdots$ &$\cdots$&$\cdots$ &0.096 & &0.021 &$\cdots$&$\cdots$& &0.173 & & & & &$\cdots$&$\cdots$& &$\cdots$ \\
		\hline
		${5/2}^{-}$    &6.857 &$\ast$ & & & & & & & & & & & & & & & & & \\
		${3/2}^{-}$    &6.843 &0.171 &0.022 &0.061 & &0.005 & & & & & &0.071 & & 0.004& & & & & \\
		&6.827 &0.019 &0.035 &0.190 & & 0.004& & & & & &0.109 & &$<10^{-4}$ & & & & & \\
		&6.653 &$\cdots$&0.008 &$\cdots$& &0.015 & & & & & &0.156 & & 0.519& & & & & \\
		&6.621 &$\cdots$&0.116 &$\cdots$& &0.003 & & & & & &0.155 & &0.006 & & & & & \\
		&6.360 &$\cdots$ &$\cdots$&$\cdots$& &0.201 & & & & & &$\cdots$& &$\cdots$& & & & & \\
		${1/2}^{-}$    &6.824 &0.066 &0.059 &0.094 &0.006 & &$<10^{-4}$ & & & & &0.197 &$<10^{-4}$ &0.005 &0.002 & & & & \\
		&6.705 &$\cdots$&$<10^{-4}$ &$\cdots$&$<10^{-4}$ & &0.014 & & & & &$<10^{-4}$ &0.040 &0.310 &0.065 & & & & \\
		&6.637 &$\cdots$&0.004 &$\cdots$&0.020 & &0.001 & & & & &0.150 &0.005 &0.118 &0.396 & & & & \\
		&6.621 &$\cdots$&0.116 &$\cdots$&0.003& &$<10^{-4}$ & & & & & 0.144&$<10^{-4}$ &0.003 & 0.009& & & & \\
		&6.475 &$\cdots$ &$\cdots$&$\cdots$&0.002 & & 0.050& & & & &$\cdots$ &0.387 &0.084 &0.056 & & & & \\
		&6.325 &$\cdots$& $\cdots$& $\cdots$&0.188 & & $<10^{-4}$& & & & &$\cdots$ &0.002 &$\cdots$&$\cdots$ & & & & \\
		&6.107 &$\cdots$&$\cdots$&$\cdots$&$<10^{-4}$ & & 0.133& & & &  &$\cdots$&$\cdots$&$\cdots$&$\cdots$& & & & \\
		\bottomrule[1.0pt]
	\end{tabular}
\end{table*}

\renewcommand{\tabcolsep}{0.11cm}
\renewcommand{\arraystretch}{0.7}
\begin{table*}[!htbp]
	\caption{The bag radius $R_{0}$ is measured in units of $\text{GeV}^{-1}$, the mass spectrum is expressed in GeV, and the magnetic moment is given in $\mu_N$, $\delta m$ is the mass calculated relative to the corresponding threshold energy.}
	\label{1tab:nnscn}
	\begin{tabular}{cc|cc|cccccccccccc}
		\bottomrule[1.0pt]
		\multirow{2}{*}{F-symmetry} &\multirow{2}{*}{$J^{P}$} 
		&\multicolumn{2}{l|}{$P_{nnsc\bar{n}}\quad R_{0}=5.90\,$GeV$^{-1}$}
		&\multicolumn{12}{l}{$\delta m=M-M_{Threshold}$} \\
		& &$M$ &$\mu$ &$\Xi_{c}^{\ast}\omega$ &$\Xi_{c}\omega$ &$\Xi_{c}^{\ast}\pi$ &$\Xi_{c}\pi$ 
		&$\Sigma_{c}^{\ast}K^{\ast}$ &$\Sigma_{c}K^{\ast}$ &$\Sigma_{c}^{\ast}K$ &$\Sigma_{c}K$  
		&$\Lambda_{c}K^{\ast}$ &$\Lambda_{c}K$ &$\Lambda D^{\ast}$ &$\Lambda D$ \\ \hline
		$\{nn\}$
		&${5/2}^{-}$    &3.500  &5.09, 1.87, -1.35, -4.57 &0.071 & & & &0.088 & & & & & & & \\
		&               &3.476  &5.09, 1.87, -1.35, -4.57 &0.047 & & & &0.064 & & & & & & & \\
		&${3/2}^{-}$    &3.567 &2.76, 1.46, -0.43, -1.73 &0.138 &0.315 &0.781 & &0.155 &0.219 &0.553& &0.387 & &0.442 & \\
		&               &3.493&0.44, 1.51, -0.04, -2.13 &0.064 &0.241 &0.707 & &0.081&0.145 &0.479 & &0.313 & &0.368 & \\
		&               &3.453  &3.90, 0.99, -1.66, -4.58 &0.024 &0.201 &0.667 & &0.041 &0.105 &0.439 & &0.273 & &0.328 & \\
		&               &3.410  &2.93, 0.28, -1.82, -4.47 &-0.019 &0.158 &0.624 & &-0.002 &0.062 &0.396 & &0.230 & &0.285 & \\
		&               &3.301  &3.92, 2.14, -0.49, -2.27 &-0.128 &0.049 &0.515 & &-0.111 &-0.047 &0.287 & &0.121 & &0.176 & \\
		&               &3.286  &4.43, 1.98, -0.24, -2.68 &-0.143 &0.034 &0.500 & &-0.126 &-0.062 &0.272 & &0.106 & &0.161 & \\
		&               &3.076  &3.33, 1.98, -0.30, -1.65 &-0.353 &-0.176 &0.290 & &-0.336 &-0.272 &0.062 & &-0.104 & &-0.049 & \\
		&${1/2}^{-}$    &3.677  &0.78, 0.91, 0.23, 0.36 &0.248 &0.425 & &1.068 &0.265 &0.329 & &0.727 &0.497 &0.895 &0.552 &0.693 \\
		&               &3.576  &1.04, 0.23, -0.49, -1.31 &0.147 &0.324 & &0.967 &0.164 &0.228 & &0.626 &0.396 &0.794 &0.451 &0.592 \\
		&               &3.466  &0.92, 0.25, -0.66, -1.33 &0.037 &0.214 & &0.857 &0.054 &0.118 & &0.516 &0.286 &0.684 &0.341 &0.482 \\
		&               &3.354  &-0.02, -0.34, 0.16, -0.16 &-0.075 &0.102 & &0.745 &-0.058 &0.006 & &0.404 &0.174 &0.572 &0.229 &0.370 \\
		&               &3.293  &1.82, 0.59, -0.70, -1.93 &-0.136 &0.041 & &0.684 &-0.119 &-0.055 & &0.343 &0.113 &0.511 &0.168 &0.309 \\
		&               &3.251  &2.65, 1.00, -0.50, -2.14 &-0.178 &-0.001 & &0.642 &-0.161 &-0.097 & &0.301 &0.071 &0.469 &0.126 &0.267 \\
		&               &3.049  &0.00, 0.25, 0.23, 0.48 &-0.380 &-0.203 & &0.440 &-0.363 &-0.299 & &0.099 &-0.131 &0.267 &-0.076 &0.065 \\
		&               &2.969  &2.47, 1.39, -0.42, -1.50 &-0.460 &-0.283 & &0.360 &-0.443 &-0.379 & &0.019 &-0.211 &0.187 &-0.156 &-0.015 \\
		\hline
		$[nn]$
		&${5/2}^{-}$    &3.476  &1.87, -1.35 &0.047 & & & &0.064 & & & & & & & \\
		&${3/2}^{-}$    &3.459  &0.97, -0.14 &0.030 &0.207 &0.673 & &0.047 &0.111 &0.445 & &0.279 & &0.334 & \\
		&               &3.407  &1.68, -1.53 &-0.022 &0.155 &0.621 & &-0.005 &0.059 &0.393 & &0.227 & &0.282 & \\
		&               &3.255  &0.75, -2.19 &-0.174 &0.003 &0.469 & &-0.157 &-0.093 &0.241 & &0.075 & &0.130 & \\
		&               &3.225  &1.54, -1.64 &-0.204 &-0.027 &0.439 & &-0.187 &-0.123 &0.211 & &0.045 & &0.100 & \\
		&               &2.985  &0.35, -0.15 &-0.444 &-0.267 &0.199 & &-0.427 &-0.363 &-0.029 & &-0.195 & &-0.140 & \\
		&${1/2}^{-}$    &3.426  &0.40, -0.16 &-0.003 &0.174 & &0.817 &0.014 &0.078 & &0.476 &0.246 &0.644 &0.301 &0.442 \\
		&               &3.299  &1.02, -0.04 &-0.130 &0.047 & &0.690 &-0.113 &-0.049 & &0.349 &0.119 &0.517 &0.174 &0.315 \\
		&               &3.227  &0.35, -1.60 &-0.202 &-0.025 & &0.618 &-0.185 &-0.121 & &0.277 &0.047 &0.445 &0.102 &0.243 \\
		&               &3.219  &0.24, -1.79 &-0.210 &-0.033 & &0.610 &-0.193 &-0.129 & &0.269 &0.039 &0.437 &0.094 &0.235 \\
		&               &3.069  &0.24, 0.91 &-0.360 &-0.183 & &0.460 &-0.343 &-0.279 & &0.119 &-0.111 &0.287 &-0.056 &0.085 \\
		&               &2.884  &-0.32, -0.69 &-0.545 &-0.368 & &0.275 &-0.528 &-0.464 & &-0.066 &-0.296 &0.102 &-0.241 &-0.100 \\
		&               &2.713  &0.48, 0.40 &-0.716 &-0.539 & &0.104 &-0.699 &-0.635 & &-0.237 &-0.467 &-0.069 &-0.412 &-0.271 \\
		\bottomrule[1.0pt]
	\end{tabular}
\end{table*}

For the mass spectra, magnetic moment, bag radius, and threshold comparison of the $nnsc\bar{n}$ system, we present the results in Table \ref{1tab:nnscn}. The results show that the average bag radius of $nnsc\bar{n}$ is nearly consistent with that of $nnnc\bar{n}$, approximately $6.00 \, \text{GeV}^{-1}$($1.18\, \text{fm}$). The mass range is between $2.7\, \text{GeV}$ and $3.7 \, \text{GeV}$, with all states lying above the lowest threshold. Additionally, most states have masses approximately $500\, \text{MeV}$ above the ground state baryon $\Xi_{c}(1/2^+, 2467)$. Here, the difference between $\{\overline{P_{nnsQ\bar{n}}}\}$ and $[\overline{P_{nnsc\bar{n}}}]$ is approximately 210 MeV, which roughly aligns with the relationship given in Eq. (\ref{2tippair}). Coincidentally, a negative parity mirror pentaquark with a spin-parity of $J^P = 1/2^{-}$ and a mass of $3302 \, \text{MeV}$ was predicted in reference \cite{PhysRevD.104.034009}. This value is very close to our result of $P_{nnsc\bar{n}}(1/2^{-}, 3299)$.

In Table \ref{2tab:nnscn}, we analyze the strong decay stability of the $nnsc\bar{n}$ system across different representations. Similar to other singly-strange pentaquark systems, most states are strongly decay unstable; however, there are several states that merit attention, such as $P_{nnsc\bar{n}}(1/2^{-}, 2713)$, $P_{nnsc\bar{n}}(1/2^{-}, 2884)$,\\ $P_{nnsc\bar{n}}(1/2^{-}, 2969)$, and $P_{nnsc\bar{n}}(3/2^{-}, 3076)$. The maximum decay widths of these states are all less than $150\, \text{MeV}$.

\renewcommand{\tabcolsep}{0.04cm}
\renewcommand{\arraystretch}{0.8}
\begin{table*}[!htbp]
	\caption{The characteristic width $|c_i|^2 \cdot k$ is measured in units of GeV, $\ast$ denotes scattering states, and $\cdots$ indicates threshold suppression.}
	\label{2tab:nnscn}
	\begin{tabular}{cc|cccccc|cccccc|cccccc}
		\bottomrule[1.0pt]
			\multicolumn{2}{c|}{} & \multicolumn{6}{c|}{$nsc\otimes n\bar{n}$} &\multicolumn{6}{c|}{$nnc\otimes s\bar{n}$} & \multicolumn{6}{c}{$nns\otimes c\bar{n}$} \\
		$J^{P}$&$M$ &$\Xi_{c}^{\ast}\omega/\rho$ &$\Xi_{c}\omega/\rho$&$\Xi_{c}^{\prime}\omega/\rho$ &$\Xi_{c}^{\prime}\pi$&$\Xi_{c}^{\ast}\pi$ &$\Xi_{c}\pi$ 
		&$\Sigma_{c}^{\ast}K^{\ast}$ &$\Sigma_{c}K^{\ast}$ &$\Sigma_{c}^{\ast}K$ &$\Sigma_{c}K$  
		&$\Lambda_{c}K^{\ast}$ &$\Lambda_{c}K$ &$\Lambda D^{\ast}$ &$\Lambda D$&$\Sigma D^{\ast}$&$\Sigma D$&$\Sigma^{\ast} D$&$\Sigma^{\ast} D^{\ast}$ \\ \hline
		${5/2}^{-}$
		&3.500 & & & & & & &$\ast$ & & & & & & & & & & & \\
		&3.476 &$\ast$ & & & & & & & & & & & & & & & & & \\
		${3/2}^{-}$
		&3.567 &0.072 &$<10^{-4}$ &$<10^{-4}$ & &0.002 & &0.108 &0.031 &$<10^{-4}$ & & & & & &0.001 & &0.097 &0.238 \\
		&3.493 &0.061 &0.072 &$<10^{-4}$ & &$<10^{-4}$ & &0.189 &$<10^{-4}$ &0.026 & & & & & &0.016 & &$<10^{-4}$ &0.001 \\
		&3.453 &0.008 &$<10^{-4}$ &0.004 & &0.024 & &0.032 &0.053 &0.035 & & & & & &0.001 & &0.188 &0.151 \\
		&3.410 &... &0.026 &0.056 & &0.001 & &$\cdots$&0.201 &$<10^{-4}$ & & & & & &0.004 & &0.008 &0.003 \\
		&3.301 &$\cdots$&0.012 &$\cdots$& &0.048 & &$\cdots$&$\cdots$&0.172 & & & & & &0.026 & &0.122 &$\cdots$\\
		&3.286 &$\cdots$&0.031 &$\cdots$& &0.021 & &$\cdots$&$\cdots$&0.003 & & & & & &0.267 & &0.017 &$\cdots$\\
		&3.076 &$\cdots$&$\cdots$&$\cdots$& &0.097 & &$\cdots$&$\cdots$ &0.135 & & & & & &$\cdots$& &$\cdots$&$\cdots$\\
		${1/2}^{-}$
		&3.677 &0.212 &$<10^{-4}$ &0.012 &$<10^{-4}$ & &$<10^{-4}$ &0.236 &0.012 & &$<10^{-4}$ & & & & &$<10^{-4}$ &$<10^{-4}$ & &0.001 \\
		&3.576 &$<10^{-4}$ &0.001 &0.147 &$<10^{-4}$ & &$<10^{-4}$ &0.002 &0.149 & &$<10^{-4}$ & & & & &0.001 &$<10^{-4}$ & &0.001 \\
		&3.466 &0.020 &0.112 &0.014 &$<10^{-4}$ & &0.002 &0.091 &0.073 & &0.010 & & & & &0.044 &0.010 & &0.002 \\
		&3.354 &$\cdots$&0.007 &$\cdots$&$<10^{-4}$ & &0.012 &$\cdots$ &0.030 & &0.005 & & & & &0.234 &0.037 & &$\cdots$\\
		&3.293 &$\cdots$&0.003 &$\cdots$&0.098 & &$<10^{-4}$ &$\cdots$&$\cdots$& &0.202 & & & & &$<10^{-4}$ &0.003 & &$\cdots$\\
		&3.251 &$\cdots$&$\cdots$&$\cdots$&0.017 & &0.014 &$\cdots$&$\cdots$ & &0.016 & & & & &0.053 &0.286 & &$\cdots$\\
		&3.049 &$\cdots$&$\cdots$&$\cdots$&0.016 & &0.172 &$\cdots$&$\cdots$ & &0.025 & & & & &$\cdots$&$\cdots$& &$\cdots$\\
		&2.969 &$\cdots$&$\cdots$&$\cdots$&0.065 & &0.018 &$\cdots$&$\cdots$ & &0.066 & & & & &$\cdots$&$\cdots$& &$\cdots$\\
		\hline
		${5/2}^{-}$
		&3.476 &$\ast$ & & & & & & & & & & & & & & & & & \\
		${3/2}^{-}$
		&3.459 &0.097 &0.026 &0.008 & &0.008 & & & & & &0.082 & &0.014 & & & & & \\
		&3.407 &$\cdots$&0.019 &0.135 & &0.001 & & & & & &0.052 & &0.006 & & & & & \\
		&3.255 &$\cdots$&0.003 &$\cdots$& &0.024 & & & & & &0.086 & &0.353 & & & & & \\
		&3.225 &$\cdots$&$\cdots$&$\cdots$& &0.002 & & & & & &0.106 & &$<10^{-4}$ & & & & & \\
		&2.985 &$\cdots$&$\cdots$&$\cdots$& &0.159 & & & & & &$\cdots$& &$\cdots$& & & & & \\
		${1/2}^{-}$
		&3.426 &$\cdots$&0.046 &0.041 &0.003 & &0.001 & & & & &0.157 &0.003 &0.037 &0.004 & & & & \\
		&3.299 &$\cdots$&0.001 &$\cdots$&0.001 & &0.009 & & & & &0.006 &0.030 &0.216 &0.033 & & & & \\
		&3.227 &$\cdots$&$\cdots$&$\cdots$&0.006 & &$<10^{-4}$ & & & & &0.056 &0.006 &0.009 &0.038 & & & & \\
		&3.219 &$\cdots$&$\cdots$&$\cdots$&0.017 & &0.006 & & & & &0.099 &0.026 &0.075 &0.250 & & & & \\
		&3.069 &$\cdots$&$\cdots$&$\cdots$&0.013 & &0.039 & & & & &... &0.283 &$\cdots$&0.071 & & & & \\
		&2.884 &$\cdots$&$\cdots$&$\cdots$&0.130 & &$<10^{-4}$ & & & & &$\cdots$&0.007 &$\cdots$&$\cdots$& & & & \\
		&2.713 &$\cdots$&$\cdots$&$\cdots$&$\cdots$& &0.093 & & & & &$\cdots$&$\cdots$&$\cdots$&$\cdots$& & & & \\
		\bottomrule[1.0pt]
	\end{tabular}
\end{table*}

\subsection{$ssqQ\bar{q}$ Systems}

\renewcommand{\tabcolsep}{0.16cm}
\renewcommand{\arraystretch}{0.7}
\begin{table*}[!htbp]
	\caption{The bag radius $R_{0}$ is measured in units of $\text{GeV}^{-1}$, the mass spectrum is expressed in GeV, and the magnetic moment is given in $\mu_N$, $\delta m$ is the mass calculated relative to the corresponding threshold energy.}
	\label{1tab:ssnbn}
	\begin{tabular}{c|cc|cccccccccccc}
		\bottomrule[1.0pt]
		\multirow{2}{*}{$J^{P}$} 
		&\multicolumn{2}{l|}{$P_{ssnb\bar{n}}\quad R_{0}=5.92\,$GeV$^{-1}$}
		&\multicolumn{12}{l}{$\delta m=M-M_{Threshold}$} \\
		&$M$ &$\mu$ &$\Omega_{b}^{\ast}\omega$ &$\Omega_{b}\omega$ &$\Omega_{b}^{\ast}\pi$ &$\Omega_{b}\pi$
		&$\Xi^{\ast}B^{\ast}$ &$\Xi^{\ast}B$ &$\Xi B^{\ast}$ &$\Xi B$
		&$\Xi_{b}^{\ast}K^{\ast}$ &$\Xi_{b}K^{\ast}$ &$\Xi_{b}^{\ast}K$ &$\Xi_{b}K$ \\ \hline
		${5/2}^{-}$     &7.030  &1.53, -1.64, -4.80 &0.135 & & & &0.172 & & & &0.182 & & & \\
		&6.982  &1.53, -1.64, -4.80 &0.087 & & & &0.124 & & & &0.134 & & & \\
		${3/2}^{-}$     &7.102  &0.23, -0.85, -1.93 &0.207 &0.273 &0.850 & &0.244 &0.289 &0.459 & &0.254 &0.414 &0.652 & \\
		&7.008  &1.63, -1.29, -4.19 &0.113 &0.179 &0.756 & &0.150 &0.195 &0.365 & &0.160 &0.320 &0.558 & \\
		&6.976  &0.54, -1.07, -2.69 &0.081 &0.147 &0.724 & &0.118 &0.163 &0.333 & &0.128 &0.288 &0.526 & \\
		&6.959  &2.24, -0.88, -3.99 &0.064 &0.130 &0.707 & &0.101 &0.146 &0.316 & &0.111 &0.271 &0.509 & \\
		&6.858  &1.27, -0.58, -2.43 &-0.037 &0.029 &0.606 & &0.000 &0.045 &0.215 & &0.010 &0.170 &0.408 & \\
		&6.833  &-0.56, -1.67, -2.77 &-0.062 &0.004 &0.581 & &-0.025 &0.020 &0.190 & &-0.015 &0.145 &0.383 & \\
		&6.530  &-1.07, -1.15, -1.82 &-0.365 &-0.299 &0.278 & &-0.328 &-0.283 &-0.113 & &-0.318 &-0.158 &0.080 & \\
		${1/2}^{-}$     &7.159  &-0.17, -0.06, 0.06 &0.264 &0.330 & &0.973 &0.301 & &0.516 &0.561 &0.311 &0.471 & &0.869 \\
		&7.110  &0.26, -0.49, -1.24 &0.215 &0.281 & &0.924 &0.252 & &0.467 &0.512 &0.262 &0.422 & &0.820 \\
		&6.963  &0.81, -0.34, -1.48 &0.068 &0.134 & &0.777 &0.105 & &0.320 &0.365 &0.115 &0.275 & &0.673 \\
		&6.872  &0.44, 0.03, -0.38 &-0.023 &0.043 & &0.686 &0.014 & &0.229 &0.274 &0.024 &0.184 & &0.582 \\
		&6.849  &1.02, -0.30, -1.62 &-0.046 &0.020 & &0.663 &-0.009 & &0.206 &0.251 &0.001 &0.161 & &0.559 \\
		&6.816  &-0.48, -1.06, -1.65 &-0.079 &-0.013 & &0.630 &-0.042 & &0.173 &0.218 &-0.032 &0.128 & &0.526 \\
		&6.657  &-0.17, -0.08, 0.00 &-0.238 &-0.172 & &0.471 &-0.201 & &0.014 &0.059 &-0.191 &-0.031 & &0.367 \\
		&6.500  &-0.70, -0.92, -1.13 &-0.395 &-0.329 & &0.314 &-0.358 & &-0.143 &-0.098 &-0.348 &-0.188 & &0.210 \\
		\bottomrule[1.0pt]
	\end{tabular}
\end{table*}

\renewcommand{\tabcolsep}{0.17cm}
\renewcommand{\arraystretch}{0.8}
\begin{table*}[!htbp]
	\caption{The characteristic width $|c_i|^2 \cdot k$ is measured in units of GeV, $\ast$ denotes scattering states, and $\cdots$ indicates threshold suppression.}
	\label{2tab:ssnbn}
	\begin{tabular}{cc|cccc|cccccc|cccc}
		\bottomrule[1.0pt]
		\multicolumn{2}{c|}{} & \multicolumn{4}{c|}{$ssb\otimes n\bar{n}$} &\multicolumn{6}{c|}{$nsb\otimes s\bar{n}$} & \multicolumn{4}{c}{$nss\otimes b\bar{n}$} \\
		$J^{P}$&$M$ &$\Omega_{b}^{\ast}\omega/\rho$ &$\Omega_{b}^{\ast}\pi$&$\Omega_{b}\omega/\rho$ &$\Omega_{b}\pi$&$\Xi_{b}^{\ast}K^{\ast}$ &$\Xi_{b}^{\ast}K$ & $\Xi_{b}^{\prime}K^{\ast}$ &$\Xi_{b}^{\prime}K$ &$\Xi_{b}K^{\ast}$ &$\Xi_{b}K$ &$\Xi^{\ast}B^{\ast}$ &$\Xi^{\ast}B$ &$\Xi B^{\ast}$ &$\Xi B$ \\ \hline
		${5/2}^{-}$     &7.030  & & & & &$\ast$ & & & & & & & & & \\
		&6.982  &$\ast$ & & & & & & & & & & & & & \\
		${3/2}^{-}$     &7.102  &0.081 &0.001 &0.027 & &0.131 &$<10^{-4}$ &0.035 & & $<10^{-4}$& &0.362 &0.155 &0.002 & \\
		&7.008  &0.001 &0.003 &$<10^{-4}$ & &0.037 &0.003 &0.083 & &$<10^{-4}$ & &0.208 &0.368 &$<10^{-4}$ & \\
		&6.976  &0.259 &0.002 &0.020 & &0.044 &0.011 &0.002 & &0.093 & &0.010 &0.009 &0.003 & \\
		&6.959  &$<10^{-4}$ &0.002 &0.378 & &0.002 &0.001 &0.053 & &0.052 & &0.011 &0.029 &$<10^{-4}$ & \\
		&6.858  &$\cdots$&0.079 &0.005 & &0.011 &0.187 &0.006 & &0.008 & &0.007 &0.072 &0.001 & \\
		&6.833  &$\cdots$ &0.020 &0.002 &  &$\cdots$&$<10^{-4}$ &0.001 & &0.077 & &$\cdots$&0.001 &0.557 &\\
		&6.530  &$\cdots$&0.323 &$\cdots$& &$\cdots$&0.037 &$\cdots$& &$\cdots$& &$\cdots$&$\cdots$&$\cdots$&\\
		${1/2}^{-}$     &7.159  &0.219 & &0.044 &$<10^{-4}$ &0.277 & &0.046 &$<10^{-4}$ &$<10^{-4}$ &$<10^{-4}$ &0.053 & &$<10^{-4}$ &$<10^{-4}$ \\
		&7.110  &0.004 & &0.137 &$<10^{-4}$ &0.006 & &0.170 &$<10^{-4}$ &$<10^{-4}$ &$<10^{-4}$ &0.485 & &0.001 &0.001 \\
		&6.963  &0.089 & &0.150 &0.002 &0.021 & &0.027 &0.048 &0.158 &$<10^{-4}$ &0.001 & &0.009 &0.001 \\
		&6.872  &$\cdots$& &0.072 &$<10^{-4}$ &0.011 & &0.011 &0.031 &$<10^{-4}$ &0.019 &$<10^{-4}$ & &0.361 &0.060  \\
		&6.849  &$\cdots$& &$<10^{-4}$ &0.091 &0.001 & &0.018 &0.072 &0.006 &$<10^{-4}$ &$\cdots$& &0.018 &0.002 \\
		&6.816  &$\cdots$& &$\cdots$&0.029 &$\cdots$&$\cdots$&$\cdots$ &0.007 &0.066 &0.004 &... & &0.108 &0.449 \\
		&6.657  &$\cdots$& &$\cdots$&0.003 &$\cdots$&$\cdots$&$\cdots$ &0.005 &$\cdots$&0.257 &$\cdots$& &0.055 &0.049\\
		&6.500  &$\cdots$& &$\cdots$&0.344 &$\cdots$&$\cdots$&$\cdots$ &$<10^{-4}$ &$\cdots$&0.001 &$\cdots$& &$\cdots$&$\cdots$\\
		\bottomrule[1.0pt]
	\end{tabular}
\end{table*}

In Table \ref{1tab:ssnbn}, we present the bag radius, mass spectra, magnetic moments, and thresholds for the $ssnb\bar{n}$ system with strangeness $S = -2$. The bag radius is nearly the same as that of the $nnnb\bar{n}$ and $nnsb\bar{n}$ systems, measured at $5.9 \, \text{GeV}^{-1}$ ($1.16\, \text{fm}$). The masses range from $6.5$ to $7.1\, \text{GeV}$, with the minimum mass approximately $450\, \text{MeV}$ above the ground state $\omega_{b}(1/2^+, 6046)$. All these states lie above the lowest thresholds, with some being close to the threshold. In Table \ref{2tab:ssnbn}, We provide the characteristic decay widths $k \cdot |c_i|^2$ for the $ssnb\bar{n}$ system in various representations. Most states have maximum characteristic decay widths greater than $200\, \text{MeV}$; however, two states exhibit smaller widths. For instance, the state $P_{ssnb\bar{n}}(1/2^{-}, 6849)$ has its largest decay channel as $\Omega_{b}\pi$, with a characteristic width of $90\, \text{MeV}$, while the state $P_{ssnb\bar{n}}(3/2^{-}, 6858)$ has its largest decay channel as $\Xi_{b}^{\ast}K$, with a characteristic width of $187\, \text{MeV}$. Therefore, these two states are promising candidates for experimental identification.

\renewcommand{\tabcolsep}{0.17cm}
\renewcommand{\arraystretch}{0.8}
\begin{table*}[!htbp]
	\caption{The bag radius $R_{0}$ is measured in units of $\text{GeV}^{-1}$, the mass spectrum is expressed in GeV, and the magnetic moment is given in $\mu_N$, $\delta m$ is the mass calculated relative to the corresponding threshold energy.}
	\label{1tab:ssncn}
	\begin{tabular}{c|cc|cccccccccccc}
		\bottomrule[1.0pt]
		\multirow{2}{*}{$J^{P}$} 
		&\multicolumn{2}{l|}{$P_{ssnc\bar{n}}\quad R_{0}=5.92\,$GeV$^{-1}$}
		&\multicolumn{12}{l}{$\delta m=M-M_{Threshold}$} \\
		&$M$ &$\mu$ &$\Omega_{c}^{\ast}\omega$ &$\Omega_{c}\omega$ &$\Omega_{c}^{\ast}\pi$ &$\Omega_{c}\pi$
		&$\Xi^{\ast}D^{\ast}$ &$\Xi^{\ast}D$ &$\Xi D^{\ast}$ &$\Xi D$
		&$\Xi_{c}^{\ast}K^{\ast}$ &$\Xi_{c}K^{\ast}$ &$\Xi_{c}^{\ast}K$ &$\Xi_{c}K$ \\ \hline
		${5/2}^{-}$     &3.643  &2.17, -1.06, -4.29 &0.094 & & & &0.101 & & & &0.103 & & & \\
		&3.607  &2.17, -1.06, -4.29 &0.058 & & & &0.065 & & & &0.067 & & & \\
		${3/2}^{-}$     &3.697  &0.93, -0.31, -1.54 &0.148 &0.219 &0.791 & &0.155 &0.296 &0.370 & &0.157 &0.334 &0.555 & \\
		&3.600  &0.71, -0.87 -2.45 &0.051 &0.122 &0.694 & &0.058 &0.199 &0.273 & &0.060 &0.237 &0.458 & \\
		&3.595  &1.81, -1.15, -4.10 &0.046 &0.117 &0.689 & &0.053 &0.194 &0.268 & &0.055 &0.232 &0.453 & \\
		&3.547  &2.16, -0.99, -4.15 &-0.002 &0.069 &0.641 & &0.005 &0.146 &0.220 & &0.007 &0.184 &0.405 & \\
		&3.447  &1.49, -0.31, -2.10 &-0.102 &-0.031 &0.541 & &-0.095 &0.046 &0.120 & &-0.093 &0.084 &0.305 & \\
		&3.436  &0.04, -1.13, -2.28 &-0.113 &-0.042 &0.530 & &-0.106 &0.035 &0.109 & &-0.104 &0.073 &0.294 & \\
		&3.160  &-0.48, -0.87, -1.25 &-0.389 &-0.318 &0.254 & &-0.382 &-0.241 &-0.167 & &-0.38 &-0.203 &0.018 & \\
		${1/2}^{-}$     &3.789  &0.11, 0.21, 0.30 &0.240 &0.311 & &0.954 &0.247 & &0.462 &0.603 &0.249 &0.426 & &0.824 \\
		&3.703  &0.36, -0.35, -1.06 &0.154 &0.225 & &0.868 &0.161 & &0.376 &0.517 &0.163 &0.340 & &0.738 \\
		&3.572  &0.47, -0.43, -1.31 &0.023 &0.094 & &0.737 &0.030 & &0.245 &0.386 &0.032 &0.209 & &0.607 \\
		&3.470  &1.35, 0.47, -0.41 &-0.079 &-0.008 & &0.635 &-0.072 & &0.143 &0.284 &-0.070 &0.107 & &0.505 \\
		&3.438  &0.94, -0.47, -1.87 &-0.111 &-0.040 & &0.603 &-0.104 & &0.111 &0.252 &-0.102 &0.075 & &0.473 \\
		&3.399  &-0.55, -1.14, -1.73 &-0.150 &-0.079 & &0.564 &-0.143 & &0.072 &0.213 &-0.141 &0.036 & &0.434 \\
		&3.255  &0.16, 0.39, 0.62 &-0.294 &-0.223 & &0.420 &-0.287 & &-0.072 &0.069 &-0.285 &-0.108 & &0.290 \\
		&3.070  &-1.01, -1.16, -1.30 &-0.479 &-0.408 & &0.235 &-0.472 & &-0.257 &-0.116 &-0.470 &-0.293 & &0.105 \\
        \bottomrule[1.0pt]
	\end{tabular}
\end{table*}

\renewcommand{\tabcolsep}{0.15cm}
\renewcommand{\arraystretch}{0.8}
\begin{table*}[!htbp]
	\caption{The characteristic width $|c_i|^2 \cdot k$ is measured in units of GeV, $\ast$ denotes scattering states, and $\cdots$ indicates threshold suppression.}
	\label{2tab:ssncn}
	\begin{tabular}{cc|cccc|cccccc|cccc}
		\bottomrule[1.0pt]
		\multicolumn{2}{c|}{} & \multicolumn{4}{c|}{$ssc\otimes n\bar{n}$} &\multicolumn{6}{c|}{$nsc\otimes s\bar{n}$} & \multicolumn{4}{c}{$nss\otimes c\bar{n}$} \\
		$J^{P}$&$M$ &$\Omega_{c}^{\ast}\omega/\rho$ &$\Omega_{c}^{\ast}\pi$&$\Omega_{b}\omega/\rho$ &$\Omega_{c}\pi$&$\Xi_{c}^{\ast}K^{\ast}$ &$\Xi_{b}^{\ast}K$ & $\Xi_{c}^{\prime}K^{\ast}$ &$\Xi_{c}^{\prime}K$ &$\Xi_{c}K^{\ast}$ &$\Xi_{c}K$ &$\Xi^{\ast}D^{\ast}$ &$\Xi^{\ast}D$ &$\Xi D^{\ast}$ &$\Xi D$ \\ \hline
		${5/2}^{-}$     &3.643  & & & & &$\ast$ & & & & & & & & & \\
		&3.607  &$\ast$ & & & & & & & & & & & & & \\
		${3/2}^{-}$     &3.697  &0.058 &0.001 &0.033 & &0.087  &$<10^{-4}$ &0.034 & &$<10^{-4}$ & &$<10^{-4}$ &0.031 &$<10^{-4}$ &\\
		&3.600  &0.204 &$<10^{-4}$ &0.006 & &0.010 &0.025 &$<10^{-4}$ & &0.033 & &0.004 &$<10^{-4}$ &0.045 & \\
		&3.595  &0.005 &0.025 &0.001 & &0.056 &0.014 &0.032 & &0.004 & &0.002 &0.027 &0.006 & \\
		&3.547  &$\cdots$&0.002 &0.230 & &$<10^{-4}$ &$<10^{-4}$ &0.048 & &0.024 & &$<10^{-4}$ &0.038 &0.035 & \\
		&3.447  &$\cdots$&0.034 &$\cdots$& &$\cdots$&0.117 &$\cdots$& &0.005 & &$\cdots$&0.002 &0.010 & \\
		&3.436  &$\cdots$&0.030 &$\cdots$& &$\cdots$&0.010 &$\cdots$& &0.012 & &$\cdots$&0.003 &0.023 & \\
		&3.160  &$\cdots$&0.294 &$\cdots$& &$\cdots$&0.016 &$\cdots$ & &$\cdots$& &$\cdots$&$\cdots$&$\cdots$& \\
		${1/2}^{-}$     &3.789  &0.113 & &0.004 &$<10^{-4}$ &0.132 & &0.004 &$<10^{-4}$ &$<10^{-4}$ &$<10^{-4}$ &0.095 & &$<10^{-4}$ &$<10^{-4}$ \\
		&3.703  &0.001 & &0.082 &$<10^{-4}$ &0.002 & &0.089 &0.001 &$<10^{-4}$ &$<10^{-4}$ &0.139 & &0.001 &$<10^{-4}$ \\
		&3.572  &0.002 & &0.019 &$<10^{-4}$ &0.003 & &0.005 &0.003 &0.066 &0.002 &$<10^{-4}$ & &0.033 &0.004 \\
		&3.470  &$\cdots$& &$\cdots$&$<10^{-4}$ &$\cdots$& &$\cdots$&0.006 &0.001 &0.009 &... & &0.113 &0.015 \\
		&3.438  &$\cdots$& &$\cdots$&0.056 &$\cdots$& &$\cdots$&0.094 &0.001 &$<10^{-4}$ &$\cdots$& &0.003 &0.002 \\
		&3.399  &$\cdots$& &$\cdots$&0.019 &$\cdots$& &$\cdots$&$<10^{-4}$ &0.022 &$<10^{-4}$ &$\cdots$& & 0.023&0.185 \\
		&3.255  &$\cdots$& &$\cdots$&0.010 &$\cdots$& &$\cdots$&0.001 &$\cdots$&0.100 &$\cdots$& &$\cdots$&0.024 \\
		&3.070  &$\cdots$& &$\cdots$&0.082 &$\cdots$& &$\cdots$&$\cdots$ &$\cdots$ &$<10^{-4}$ &$\cdots$ & &$\cdots$&$\cdots$\\
		\bottomrule[1.0pt]
	\end{tabular}
\end{table*}

In Tables \ref{1tab:ssncn} and \ref{2tab:ssncn}, we present the mass spectra of the $ssnc\bar{n}$ system and analyze its strong decay stability. The masses of all states in the $ssnc\bar{n}$
system are located above the lowest threshold, with most states exceeding the ground state baryon $\Omega_{c}(1/2^+, 2695)$ by about 500 MeV. The characteristic widths provided in Table \ref{2tab:ssncn} indicate that the strong decay stability of the $ssnc\bar{n}$ system is generally higher than that of other flavor combinations. For example, the states $P_{ssnc\bar{n}}(3/2^{-}, 3697)$, $P_{ssnc\bar{n}}(3/2^{-}, 3595)$, $P_{ssnc\bar{n}}(3/2^{-}, 3447)$,\\ $P_{ssnc\bar{n}}(3/2^{-}, 3447)$, $P_{ssnc\bar{n}}(3/2^{-}, 3436)$, $P_{ssnc\bar{n}}(1/2^{-}, 3789)$,\\ $P_{ssnc\bar{n}}(1/2^{-}, 3572)$, $P_{ssnc\bar{n}}(1/2^{-}, 3470)$, $P_{ssnc\bar{n}}(1/2^{-}, 3438)$,\\ $P_{ssnc\bar{n}}(1/2^{-}, 3255)$, and $P_{ssnc\bar{n}}(1/2^{-}, 3070)$
all have characteristic decay widths less than $100\, \text{MeV}$.

Based on the results for singly heavy pentaquark states presented in Section \ref{sec:hadrons}, we provide some qualitative discussions, focusing mainly on two points. \newline

I.The threshold and strong decay characteristics of singly heavy pentaquarks. \newline

II.The mapping relationship between the singly heavy pentaquark $nnnQ\bar{n}$ and the singly heavy baryon $nnQ$. \newline

For the singly heavy pentaquark systems we have calculated, the bag radius generally falls between 1.16 and $1.19\, \text{fm}$. This scale is smaller than the contact radius of two $B$ mesons. However, the confinement radius for the $nnnQ\bar{n}$ state is slightly larger than the bag's binding radius and is close to the string breaking range predicted by lattice QCD. Therefore, under a compact picture, the singly heavy pentaquark states approach the limits of color interactions and may have the potential for multiquark resonances. Moreover, the masses of singly heavy pentaquarks are typically above the minimum threshold for baryon-meson pairs, indicating that these states should possess at least one strong decay channel. Additionally, the analysis of two-body strong decays shows that most singly heavy pentaquarks are strongly decay unstable; however, some specific states have characteristic widths for their main decay channels in the range of $100\textendash200\, \text{fm}$, making them promising candidates for experimental detection.

For the singly heavy quark system $nnnQ\bar{n}$, its mass spectrum is typically about $500\, \text{fm}$ higher than that of the corresponding positive-parity mirror baryon $nnQ$ ground state. This is consistent with the results estimated by chiral methods, which suggest that baryons can produce $n\bar{n}$ states at energies exceeding 500 MeV above the ground state baryon. Furthermore, our results closely align with the predictions for the two negative-parity pentaquark states found in Ref. \cite{PhysRevD.104.034009}. Additionally, we can extend the mass relations between isospin partners in singly heavy baryons, as shown in Eqs. (\ref{quark pair}) and (\ref{squark pair}), to singly heavy pentaquark states and find that our results approximately satisfy these relationships.

\section{Summary}
\label{sec:summary}

In this study, we conducted a systematic analysis of singly heavy pentaquarks with strangeness numbers $S = 0, -1, -2$ within the MIT bag framework. We observed that the bag radius for singly heavy pentaquark systems lies in the range of $1.16 \textendash 1.18 \, \text{fm}$. Within this interval, the stability of the bag is relatively weak, corresponding to a potential resonance scale. Lattice QCD suggests that the QCD vacuum becomes unstable at scales between $1.17\textendash1.29 \, \text{fm}$, and the bag radius of singly heavy pentaquarks is precisely at a sensitive scale. Furthermore, inspired by singly flavor baryons and their excited states, the mirror pentaquark states of singly flavor baryons hold greater potential for experimental discovery.

We calculated the masses of singly heavy pentaquark systems based on the bag model, taking into account the CMI between quarks and the color-electric interactions between strange quarks and heavy quarks. For the states $nnnb\bar{n}$, $nnnc\bar{n}$, $nnsb\bar{n}$, $nnsc\bar{n}$, $ssnb\bar{n}$, and $ssnc\bar{n}$, the corresponding mass ranges are $6.2 \textendash6.9 \, \text{GeV}$, $2.7\textendash3.6\, \text{GeV}$, $2.7\textendash3.6 \, \text{GeV}$, $6.5\textendash7.1\, \text{GeV}$, and $3.1\textendash3.7 \, \text{GeV}$, respectively. Almost all singly heavy pentaquark masses are above $500\, \text{MeV}$ compared to the masses of their corresponding ground state mirror baryons, which is consistent with the lower bound estimates for the generation of $n\bar{n}$ within baryons provided by chiral methods. Our mass calculations also include two singly heavy pentaquark states estimated by chiral methods, specifically $\Lambda_{c}(1/2^-, 3529)$ and $\Xi_{c}(1/2^-, 3302)$, as well as our results for $P_{nnnc\bar{n}}(1/2^-, 3569)$ and $P_{nnsc\bar{n}}(1/2^-, 3299)$.

For singly heavy pentaquarks, we can establish a mass relation between singly heavy baryons and singly heavy pentaquark systems by analyzing the flavor symmetry of the light quarks and the interactions they satisfy, as shown in Eqs. (\ref{quark pair}) and (\ref{2tippair}). Our calculations for the pentaquark states also approximately conform to this hypothesis, indicating that our results are relatively reliable.

We analyzed the stability of two-body strong decays of singly flavor pentaquarks based on various final state representations. Generally, singly heavy pentaquark states are found to be strongly decay unstable, which aligns with our assessment that these states may possess strong resonance characteristics. However, for certain states, the decay channels are limited due to threshold suppression, resulting in characteristic widths smaller than $150\, \text{MeV}$.

Both the bag model and lattice QCD provide a picture of color confinement and are similar in their curve shapes, with their respective limits for color interactions being quite close. The bag model, based on the hadronic level, can extend this limit to multiquark states. For compact singly heavy pentaquark systems, their confinement radius falls within the limit set by lattice QCD, indicating the potential for strong resonance properties. Finally, we look forward to further validation from lattice QCD and support from experimental results.

\medskip
\textbf{ACKNOWLEDGMENTS}

We thanks Wen-Yuan Wang and Ming-Zhu Liu for usefull discussings. 
The work is supported by the National Natural Science Foundation of China (Grants No. 12475026) and the Natural Science Foundation of Gansu Province\\ (No.25JRRA799).

\medskip

\section*{Appendix A: Color and Spin Wavefunctions}

For the pentaquark state, we can view it as a substructure consisting of three quarks and another substructure formed by a quark-antiquark pair. The color representation of the five-quark state can be expressed as the color direct product of these two substructures, and the group algebra can be represented by the following equation:
\begin{equation}
	(3\otimes3\otimes3)\otimes(3\otimes\bar{3})=(1\oplus8\oplus8\oplus10)\otimes(1\oplus8).
\end{equation}

In the right-hand side of the equation, two color representations $8 \otimes 8$ appear, which correspond to the color singlet represented by $1 \otimes 1$. Therefore, there are a total of three color wave functions for the pentaquarks. Here, the wave functions corresponding to the two color representations $8 \otimes 8$ are denoted as $\phi_{1}^{P}$ and $\phi_{2}^{P}$, while the wave function that describes the color singlet is represented by $\phi_{3}^{P}$, which is explicitly expressed as follows:

\begin{equation}\label{8b1}
	\begin{aligned}
		\phi_{1}^{P} &=\frac{1}{4\sqrt{3}}\Big[(2bbgr-2bbrg+gbrb-gbbr+bgrb-bgbr\\
		&-rbgb+rbbg-brgb+brbg)\bar{b}+(2rrbg-2rrgb\\
		&+rgrb-rgbr+grrb-grbr+rbgr-rbrg+brgr\\
		&-brrg)\bar{r}+(2ggrb-2ggbr-rggb+rgbg-grgb\\
		&+grbg+gbgr-gbrg+bggr-bgrg)\bar{g}\Big],
	\end{aligned}
\end{equation}

\begin{equation}\label{8b2}
	\begin{aligned}
		\phi_{2}^{P} &=\frac{1}{12}\Big[(3bgbr-3gbbr-3brbg+3rbbg-rbgb-2rgbb \\
		&+2grbb+brgb+gbrb-bgrb)\bar{b}+(3grrb-3rgrb \\
		&-3brrg+3rbrg-rbgr-2gbrr+2bgrr-grbr  \\
		&+rgbr+brgr)\bar{r}+(3grgb-3rggb+3bggr-3gbgr \\
		&-grbg+rgbg+2rbgg-2brgg+gbrg-bgrg)\bar{g}\Big],    
	\end{aligned}
\end{equation}

\begin{equation}\label{8b3}
	\begin{aligned}
		\phi_{3}^{P} &=\frac{1}{3\sqrt{2}}\Big[(grbb-rgbb+rbgb-brgb+bgrb-gbrb)\bar{b}\\
		&+(grbr-rgbr+rbgr-brgr+bgrr-gbrr)\bar{r} \\
		&+(grbg-rgbg+rbgg-brgg+bgrg-gbrg)\bar{g}\Big].    
	\end{aligned}
\end{equation}

For the spin basis, it is still considered as a spin summation of a three-quark subsystem and a quark-antiquark pair subsystem:

\begin{equation}\label{Spin}
	\begin{aligned}
		\chi_{1}^{P}&=|[(12)_{1}3]_{3/2}(4\bar{5})_{1}\rangle_{5/2}, 
		\chi_{2}^{P}=|[(12)_{1}3]_{3/2}(4\bar{5})_{1} \rangle_{3/2},\\
		\chi_{3}^{P}&=|[(12)_{1}3]_{3/2}(4\bar{5})_{0}\rangle_{3/2}, 
		\chi_{4}^{P}=|[(12)_{1}3]_{1/2}(4\bar{5})_{1} \rangle_{3/2},\\
		\chi_{5}^{P}&=|[(12)_{0}3]_{1/2} (4\bar{5})_{1}\rangle_{3/2}, 
		\chi_{6}^{P}=|[(12)_{1}3]_{3/2} (4\bar{5})_{1}\rangle_{1/2},\\
		\chi_{7}^{P}&=|[(12)_{1}3]_{1/2} (4\bar{5})_{1}\rangle_{1/2},  
		\chi_{8}^{P}=|[(12)_{1}3]_{1/2}(4\bar{5})_{0}\rangle_{1/2},\\
		\chi_{9}^{P}&=|[(12)_{0}3]_{1/2}(4\bar{5})_{1}\rangle_{1/2}, 
		\chi_{10}^{P}=|[(12)_{0}3]_{1/2} (4\bar{5})_{0}\rangle_{1/2}.
	\end{aligned}
\end{equation}

Here, the numbers in parentheses represent the quarks, while the subscripts of each bracket correspond to the spins of the respective subsystems. The outermost subscript denotes the total spin of the pentaquark state. The spin basis vectors here take into account various combinations of the pentaquark.

The color-spin basis vectors of the pentaquark states are described by the algebra of the $\textbf{SU(3)} \otimes \textbf{SU(2)}$ groups. Therefore, there are a total of 30 color-spin basis vectors for the pentaquark states. Below, we classify them according to the total spin of the pentaquark states:

$J^{P}=5/2^{-}$
\begin{equation}\label{cs1}
	\begin{aligned}
		\phi_{1}\chi_{1}=|[(12)^{6}_{1}3]^{8}_{3/2} (4\bar{5})^{8}_{1}\rangle_{5/2},\\
		\phi_{2}\chi_{1}=|[(12)^{\bar{3}}_{1}3]^{8}_{3/2} (4\bar{5})^{8}_{1}\rangle_{5/2},\\
		\phi_{3}\chi_{1}=|[(12)^{\bar{3}}_{1}3]^{1}_{3/2} (4\bar{5})^{1}_{1}\rangle_{5/2}.
	\end{aligned}
\end{equation}

$J^{P}=3/2^{-}$

\begin{equation}\label{cs2}
	\begin{aligned}
		\phi_{1}\chi_{2}=|[(12)^{6}_{1}3]^{8}_{3/2} (4\bar{5})^{8}_{1}\rangle_{3/2},\\
		\phi_{1}\chi_{3}=|[(12)^{6}_{1}3]^{8}_{3/2} (4\bar{5})^{8}_{0}\rangle_{3/2},\\
		\phi_{1}\chi_{4}=|[(12)^{6}_{1}3]^{8}_{1/2} (4\bar{5})^{8}_{1}\rangle_{3/2},\\
		\phi_{1}\chi_{5}=|[(12)^{6}_{0}3]^{8}_{1/2} (4\bar{5})^{8}_{1}\rangle_{3/2},\\
		\phi_{2}\chi_{2}=|[(12)^{\bar{3}}_{1}3]^{8}_{3/2} (4\bar{5})^{8}_{1}\rangle_{3/2},\\
		\phi_{2}\chi_{3}=|[(12)^{\bar{3}}_{1}3]^{8}_{3/2} (4\bar{5})^{8}_{0}\rangle_{3/2},\\
		\phi_{2}\chi_{4}=|[(12)^{\bar{3}}_{1}3]^{8}_{1/2} (4\bar{5})^{8}_{1}\rangle_{3/2},\\
		\phi_{2}\chi_{5} = | [(12)^{\bar{3}}_{0}3]^{8}_{1/2} (4\bar{5})^{8}_{1} \rangle_{3/2},\\
		\phi_{3}\chi_{2} = | [(12)^{\bar{3}}_{1}3]^{1}_{3/2} (4\bar{5})^{1}_{1} \rangle_{3/2},  \\
		\phi_{3}\chi_{3} = | [(12)^{\bar{3}}_{1}3]^{1}_{3/2} (4\bar{5})^{1}_{0} \rangle_{3/2},  \\
		\phi_{3}\chi_{4} = | [(12)^{\bar{3}}_{1}3]^{1}_{1/2} (4\bar{5})^{1}_{1} \rangle_{3/2},  \\
		\phi_{3}\chi_{5} = | [(12)^{\bar{3}}_{0}3]^{1}_{1/2} (4\bar{5})^{1}_{1} \rangle_{3/2}.
	\end{aligned}
\end{equation}

\vspace{0.8cm}
$J^{P}=1/2^{-}$

\begin{equation}\label{cs3}
	\begin{aligned}
		&\phi_{1}\chi_{6} = | [(12)^{6}_{1}3]^{8}_{3/2} (4\bar{5})^{8}_{1} \rangle_{1/2},  \\
		&\phi_{1}\chi_{7} = | [(12)^{6}_{1}3]^{8}_{1/2} (4\bar{5})^{8}_{1} \rangle_{1/2},  \\
		&\phi_{1}\chi_{8} = | [(12)^{6}_{1}3]^{8}_{1/2} (4\bar{5})^{8}_{0} \rangle_{1/2},  \\
		&\phi_{1}\chi_{9} = | [(12)^{6}_{0}3]^{8}_{1/2} (4\bar{5})^{8}_{0} \rangle_{1/2},  \\
		&\phi_{1}\chi_{10} = | [(12)^{6}_{0}3]^{8}_{1/2} (4\bar{5})^{8}_{0} \rangle_{1/2},  \\    
		&\phi_{2}\chi_{6}  = | [(12)^{\bar{3}}_{1}3]^{8}_{3/2} (4\bar{5})^{8}_{1} \rangle_{1/2}, \\
		&\phi_{2}\chi_{7}  = | [(12)^{\bar{3}}_{1}3]^{8}_{1/2} (4\bar{5})^{8}_{1} \rangle_{1/2}, \\
		&\phi_{2}\chi_{8}  = | [(12)^{\bar{3}}_{1}3]^{8}_{1/2} (4\bar{5})^{8}_{0} \rangle_{1/2}, \\
		&\phi_{2}\chi_{9}  = | [(12)^{\bar{3}}_{0}3]^{8}_{1/2} (4\bar{5})^{8}_{1} \rangle_{1/2},  \\
		&\phi_{2}\chi_{10} = | [(12)^{\bar{3}}_{0}3]^{8}_{1/2} (4\bar{5})^{8}_{0} \rangle_{1/2},  \\
		&\phi_{3}\chi_{6}  = | [(12)^{\bar{3}}_{1}3]^{1}_{3/2} (4\bar{5})^{1}_{1} \rangle_{1/2}, \\
		&\phi_{3}\chi_{7}  = | [(12)^{\bar{3}}_{1}3]^{1}_{1/2} (4\bar{5})^{1}_{1} \rangle_{1/2}, \\
		&\phi_{3}\chi_{8} = | [(12)^{\bar{3}}_{1}3]^{1}_{1/2} (4\bar{5})^{1}_{0} \rangle_{1/2},\\
		&\phi_{3}\chi_{9}  = | [(12)^{\bar{3}}_{0}3]^{1}_{1/2} (4\bar{5})^{1}_{1} \rangle_{1/2},  \\
		&\phi_{3}\chi_{10} = | [(12)^{\bar{3}}_{0}3]^{1}_{1/2} (4\bar{5})^{1}_{0} \rangle_{1/2}.
	\end{aligned}
\end{equation}

The bases mentioned above do not consider the specific situations involving quark flavors. When taking flavor symmetry into account, it is essential to consider the conditions that satisfy this symmetry. For instance, in the $nnnQ\bar{n}$ system under the $nnn \otimes Q \bar{n}$
representation, the substructure $nnn$ can have two scenarios: one is the fully symmetric flavor case, and the other involves a pair of antisymmetric flavors. In the fully symmetric flavor case with $J^p = 5/2^-$, the only choice available is $\phi_{3}\chi_{1}$, because the flavor of the substructure is symmetric, and the spin is also symmetric; therefore, the color must be antisymmetric (corresponding to $\phi_{3}$). Consequently, the choice of basis must adhere to the Pauli exclusion principle. We have analyzed the symmetry corresponding to the flavor combinations for basis selection, which is summarized in Table \ref{tab:basis}.

\label{apd:basis}
\begin{table*}
	\caption{Considering the flavor configuration, the color-spin eigenbasis of the pentaquark state}
	\label{tab:basis}
	\begin{tabular}{lcc}
		\bottomrule[2pt]\bottomrule[0.2pt]
		System  &$J^{P}$ &Color-spin wave functions \\ \hline
		$nnn\otimes Q\bar{n}$ 
		& ${5/2}^{-}$ & $\phi_{3}\chi_{1}$ \\
		& ${3/2}^{-}$ & $\frac{1}{\sqrt{2}}(\phi_{1}\chi_{5}-\phi_{2}\chi_{4})$, $\phi_{3}\chi_{2}$, $\phi_{3}\chi_{3}$\\
		& ${1/2}^{-}$ & $\frac{1}{\sqrt{2}}(\phi_{1}\chi_{9}-\phi_{2}\chi_{7})$, $\frac{1}{\sqrt{2}}(\phi_{1}\chi_{10}-\phi_{2}\chi_{8})$, $\phi_{3}\chi_{6}$ \\
		& ${5/2}^{-}$ & $\frac{1}{\sqrt{2}}(\phi_{2}\chi_{1}-\phi_{1}\chi_{1})$ \\
		& ${3/2}^{-}$ & $\frac{1}{\sqrt{2}}(\phi_{2}\chi_{2}-\phi_{1}\chi_{2})$, $\frac{1}{\sqrt{2}}(\phi_{2}\chi_{3}-\phi_{1}\chi_{3})$, $\frac{1}{2}(\phi_{1}\chi_{5}+\phi_{2}\chi_{4}+\phi_{1}\chi_{4}-\phi_{2}\chi_{5})$, $\frac{1}{\sqrt{2}}(\phi_{3}\chi_{4}-\phi_{3}\chi_{5})$\\
		& ${1/2}^{-}$ & $\frac{1}{\sqrt{2}}(\phi_{2}\chi_{6}-\phi_{1}\chi_{6})$, $\frac{1}{2}(\phi_{1}\chi_{9}+\phi_{2}\chi_{7}+\phi_{1}\chi_{7}-\phi_{2}\chi_{9})$, $\frac{1}{2}(\phi_{1}\chi_{10}+\phi_{2}\chi_{8}+\phi_{1}\chi_{8}-\phi_{2}\chi_{10})$, $\frac{1}{\sqrt{2}}(\phi_{3}\chi_{7}+\phi_{3}\chi_{9})$, $\frac{1}{\sqrt{2}}(\phi_{3}\chi_{8}+\phi_{3}\chi_{10})$ \\
		$nnQ\otimes n\bar{n}$/$nnQ\otimes s\bar{n}$
		& ${5/2}^{-}$ & $\phi_{2}\chi_{1},\phi_{3}\chi_{1}$ \\
		& ${3/2}^{-}$ & $\phi_{1}\chi_{5},\phi_{2}\chi_{2},\phi_{2}\chi_{3},\phi_{2}\chi_{4},\phi_{3}\chi_{2},\phi_{3}\chi_{3},\phi_{3}\chi_{4}$ \\
		& ${1/2}^{-}$ & $\phi_{1}\chi_{9},\phi_{1}\chi_{10},\phi_{2}\chi_{6},\phi_{2}\chi_{7},\phi_{2}\chi_{8},\phi_{3}\chi_{6},\phi_{3}\chi_{7},\phi_{3}\chi_{8}$ \\
		
		& ${5/2}^{-}$ & $\phi_{1}\chi_{1}$ \\
		& ${3/2}^{-}$ & $\phi_{1}\chi_{2},\phi_{1}\chi_{3},\phi_{1}\chi_{4},\phi_{2}\chi_{5},\phi_{3}\chi_{5}$\\
		& ${1/2}^{-}$ & $\phi_{1}\chi_{6},\phi_{1}\chi_{7},\phi_{1}\chi_{8},\phi_{2}\chi_{9},\phi_{2}\chi_{10},\phi_{3}\chi_{9},\phi_{3}\chi_{10}$ \\
		
		$nsQ\otimes n\bar{n}$/$nsQ\otimes s\bar{n}$
		& ${5/2}^{-}$ & $\phi_{1}\chi_{1},\phi_{2}\chi_{1},\phi_{3}\chi_{1}$ \\
		& ${3/2}^{-}$ & $\phi_{1}\chi_{2},\phi_{1}\chi_{3},\phi_{1}\chi_{4},\phi_{1}\chi_{5},\phi_{2}\chi_{2},\phi_{2}\chi_{3},\phi_{2}\chi_{4},\phi_{2}\chi_{5},\phi_{3}\chi_{2},\phi_{3}\chi_{3},\phi_{3}\chi_{4},\phi_{3}\chi_{5}$ \\
		& ${1/2}^{-}$ & $\phi_{1}\chi_{6},\phi_{1}\chi_{7},\phi_{1}\chi_{8},\phi_{1}\chi_{9},\phi_{1}\chi_{10},\phi_{2}\chi_{6},\phi_{2}\chi_{7},\phi_{2}\chi_{8},\phi_{2}\chi_{9},\phi_{2}\chi_{10},\phi_{3}\chi_{6},\phi_{3}\chi_{7},\phi_{3}\chi_{8},\phi_{3}\chi_{9},\phi_{3}\chi_{10}$ \\
		
		$ssQ\otimes n\bar{n}$/$ssn\otimes Q\bar{n}$
		& ${5/2}^{-}$ & $\phi_{2}\chi_{1},\phi_{3}\chi_{1}$ \\
		& ${3/2}^{-}$ & $\phi_{1}\chi_{5},\phi_{2}\chi_{2},\phi_{2}\chi_{3},\phi_{2}\chi_{4},\phi_{3}\chi_{2},\phi_{3}\chi_{3},\phi_{3}\chi_{4}$ \\
		& ${1/2}^{-}$ & $\phi_{1}\chi_{9},\phi_{1}\chi_{10},\phi_{2}\chi_{6},\phi_{2}\chi_{7},\phi_{2}\chi_{8},\phi_{3}\chi_{6},\phi_{3}\chi_{7},\phi_{3}\chi_{8}$ \\
		\bottomrule[0.5pt]\bottomrule[1.5pt]
	\end{tabular}%
\end{table*}


\end{document}